\documentclass[11pt,a4paper]{article}

\usepackage{cite}

\usepackage{amsmath, amsthm,mathtools,setspace, url,amssymb,upgreek,textgreek,slashed,hyperref}

 \usepackage[mathscr]{eucal}
\usepackage{ifpdf}
\ifpdf
  \usepackage[pdftex]{graphicx}
  \usepackage{epstopdf}
\else
  \usepackage[dvips]{graphicx}
\fi
\parskip=\baselineskip

\def\TFD{{\mathrm{TFD}}}

\begin{document}
\begin{titlepage}
\begin{flushright}

\end{flushright}

\vskip 1.5in
\begin{center}
{\bf{Why Does Quantum Field Theory In Curved Spacetime Make Sense? }}\\ \vskip.5cm
{\bf {And What Happens To The Algebra of Observables In The Thermodynamic Limit?}
}

\vskip
0.5cm  { Edward Witten} \vskip 0.05in {\small{ \textit{Institute for Advanced Study}\vskip -.4cm
{\textit{Einstein Drive, Princeton, NJ 08540 USA}}}
}
\end{center}
\vskip 0.5in
\baselineskip 16pt
\begin{abstract}  This article aims to explain some of the basic facts about the questions
raised in the title, without  the technical details that are available in the literature.   We provide a gentle introduction
to some rather classical results about quantum field theory in curved spacetime and about the thermodynamic limit of quantum statistical
mechanics.   We also briefly explain that  these results have an analog in  the large $N$ limit of gauge theory.      \end{abstract}
\date{December, 2021}
\end{titlepage}
\def\SO{{\mathrm{SO}}}
\def\G{{\text{\sf G}}}
\def\frak{\mathfrak}
\def\la{\langle}
\def\ra{\rangle}
\def\Spinc{{\mathrm{Spin}}_c}
\def\g{{\mathfrak g}}
\def\cl{{\mathrm{cl}}}
\def\m{{\sf m}}
\def\veps{\varepsilon}
\def\Re{{\mathrm{Re}}}
\def\Im{{\mathrm{Im}}}
\def\SU{{\mathrm{SU}}}
\def\SL{{\mathrm{SL}}}
\def\a{{\sf a}}
\def\b{{\sf b}}

\def\M{{\mathcal M}}
\def\d{{\mathrm d}}
\def\g{{\mathfrak g}}
\def\su{{\mathfrak {su}}}
\def\CS{{\mathrm{CS}}}
\def\Z{{\Bbb Z}}
\def\cB{{\mathcal B}}
\def\zZ{{\mathcal Z}}
\def\DD{{\mathscr D}}
\def\R{{\Bbb R}}
\def\sF{{\sf F}}
\def\cS{{\mathcal S}}
\def\sB{{\sf B}}
\def\sA{{\sf A}}
\def\sD{{\mathcal D}}
\def\dD{{\mathrm D}}
\def\F{{ \mathscr F}}
\def\cF{{\mathcal F}}
\def\J{{\mathcal J}}
\def\Bbb{\mathbb}
\def\Tr{{\rm Tr}}
\def\ad{{\mathrm{ad}}}
\def\j{{\sf j}}
\def\16{{\bf 16}}
\def\1{{(1)}}
\def\bCP{{\Bbb{CP}}}
\def\2{{(2)}}
\def\3{{\bf 3}}
\def\4{{\bf 4}}
\def\free{{\mathrm{free}}}
\def\sg{{\mathrm g}}
\def\J{{\mathcal J}}
\def\i{{\mathrm i}}
\def\h{\widehat}
\def\HP{{{H\negthinspace P}}}
\def\u{u}
\def\D{D}
\def\Rf{{\eurm{R}}}
\def\sp{{\sigma}}
\def\E{{\mathcal E}}
\def\O{{\mathcal O}}
\def\OO{{\mathrm O}}
\def\PF{{\mathit{P}\negthinspace\mathit{F}}}
\def\tr{{\mathrm{tr}}}
\def\be{\begin{equation}}
\def\ee{\end{equation}}
 \def\Sp{{\mathrm{Sp}}}
  \def\PSp{{\mathrm{PSp}}}
 \def\Spin{{\mathrm{Spin}}}
 \def\SL{{\mathrm{SL}}}
 \def\SU{{\mathrm{SU}}}
 \def\SO{{\mathrm{SO}}}
 \def\PGL{{\mathrm{PGL}}}
 \def\ll{\langle\langle}
 \def\frL{{{\mathfrak L}}}
 \def\RR{{\mathcal R}}
\def\rr{\rangle\rangle}
\def\la{\langle}
\def\CP{{C\negthinspace P}}
\def\sCP{{\sf{CP}}}
\def\I{{\mathcal I}}
\def\ra{\rangle}
\def\T{{\mathcal T}}
\def\V{{\mathcal V}}
\def\bar{\overline}
\def\spinc{{\mathrm{spin}_c}}
\def\dim{{\mathrm{dim}}}
\def\v{v}
\def\Pic{{\mathrm{Pic}}}

\def\RP{{\Bbb{RP}}}
\def\C{{\mathbb C}}
\def\tilde{\widetilde}
\def\t{\widetilde}
\def\R{{\Bbb{R}}}
\def\N{{\mathcal N}}
\def\B{{\mathcal B}}
\def\H{{\mathcal H}}
\def\hat{\widehat}
\def\Pf{{\mathrm{Pf}}}
\def\bM{{\overline\M}}
\def\PSL{{\mathrm{PSL}}}
\def\PSU{{\mathrm{PSU}}}
\def\Im{{\mathrm{Im}}}
\def\Gr{{\mathrm{Gr}}}
\def\sign{{\mathrm{sign}}}

\font\tencmmib=cmmib10 \skewchar\tencmmib='177
\font\sevencmmib=cmmib7 \skewchar\sevencmmib='177
\font\fivecmmib=cmmib5 \skewchar\fivecmmib='177
\newfam\cmmibfam
\textfont\cmmibfam=\tencmmib \scriptfont\cmmibfam=\sevencmmib
\scriptscriptfont\cmmibfam=\fivecmmib
\def\cmmib#1{{\fam\cmmibfam\relax#1}}
\numberwithin{equation}{section}
\def\lmark{{\mathrm L}}
\def\neg{\negthinspace}

\def\C{{\Bbb C}}
\def\K{{\mathcal K}}
\def\cc{{\mathrm{cc}}}
\def\bB{{\bar {\mathscr B}}}
\def\HH{{\mathbb H}}
\def\P{{\mathcal P}}
\def\Q{{\mathcal Q}}
\def\NS{{\sf{NS}}}
\def\Ra{{\sf{R}}}
\def\sV{{\sf V}}
\def\CP{{\mathrm{CP}}}
\def\Hom{{\mathrm{Hom}}}
\def\Z{{\Bbb Z}}
\def\op{{\mathrm{op}}}
\def\bop{{\overline{\mathrm{op}}}}
\def\bA{\bar{\mathscr A}}
\def\A{{\mathscr A}}
\def\cA{{\mathcal A}}
\def\B{{\mathscr B}}
\def\S{{\sf S}}
\def\bar{\overline}
\def\sc{{\mathrm{sc}}}
\def\Max{{\mathrm{Max}}}
\def\CS{{\mathrm{CS}}}
\def\ga{\gamma}
\def\bg{\bar\ga}
\def\Arg{{\mathrm{Arg}}}
\def\W{{\mathcal W}}
\def\M{{\mathcal M}}
\def\bM{{\overline \M}}
\def\L{{\mathcal L}}
\def\sM{{\sf M}}
\def\gst{\mathrm{g}_{\mathrm{st}}}
\def\gstt{\widetilde{\mathrm{g}}_{\mathrm{st}}}
\def\hbbar{\pmb{\hbar}}
\def\G{{\mathcal G}}
\def\Sym{{\mathrm{Sym}}}
\def\U{{\mathcal U}}
\def\Bun{{\mathcal M}(G,C)}
\def\be{\begin{equation}}
\def\ee{\end{equation}}
\def\Diff{{\mathrm{Diff}}}
\def\diff{{\mathrm{diff}}}

\tableofcontents

\section{Introduction}\label{intro}

The first goal of the present article is to provide some intuition about a basic question: Why does quantum
field theory in a curved spacetime of Lorentz signature make sense?   
 Here we will give an informal introduction, referring the reader  to the literature for more detail.
An inevitably partial list of references on the general topic of quantum field theory in curved spacetime is
   \cite{DeWitt, FSW,FNW, BD,Fulling, Kay,BOS,WaldBook,BFV,F, HW,BF,PT,Fewster}.  For some technical detail on the specific topics that will be described   rather heuristically here, the reader  might
    start with the   book  \cite{WaldBook}.   Another useful reference is the collection of articles \cite{BF}.

We consider  always  a globally hyperbolic spacetime, which means a spacetime $M$
with a complete Cauchy hypersurface $S$ on which  initial conditions for classical or quantum fields can be formulated.   The sense in which quantum field
theory can be formulated on $M$ depends very much on whether $S$ is compact.   In a 
 closed universe, in other words if $S$ is compact,
the usual Hilbert space  formulation of quantum field theory is valid: to a quantum field theory on $M$, we can associate in a natural way a Hilbert space 
$\H$, such that the quantum dynamics of the given theory 
can be described  in the usual way in terms of operators acting on $\H$.    However, in contrast to the possibly more familiar case of
quantum field theory in Minkowski spacetime, there is generically no distinguished ground state or vacuum vector in $\H$, since there is no natural ``energy''
that we should try to minimize.   Thus, in formulating quantum field theory in curved spacetime, one should become accustomed to the idea of a naturally
defined Hilbert space that does not contain any distinguished vector.\footnote{A corollary is that in a general spacetime, there is no useful
notion of a particle: particles in quantum field theory (or quasiparticles in condensed matter physics) are defined  as
excitations of a distinguished ground state with certain asymptotic properties, so in the absence of any preferred state, and in the absence, for a general $M$, of any asymptotic region where asymptotic
properties could be defined, there is no notion of a ``particle.''}

The case that $S$ is not compact, and with no simplifying assumptions about the behavior at infinity along $S$, is quite different.   In this case, a natural
construction of a Hilbert space does not exist.   Instead, roughly speaking, one has to work with density matrices for local algebras of observables, more precisely
an algebra $\A_U$ for each bounded open set $U\subset M$. And roughly speaking, quantum field theory on $M$ describes the evolution not of a quantum state but of a density matrix.  Actually a slight generalization of the notion of a density matrix is needed, 
because in quantum field theory the algebras $\A_U$ are von Neumann algebras of Type III, not of Type I
\cite{ArakiQFT,Fred,HaagBook,Yng,WittenNotes}, as  will be explained in due course.

To be more precise,
 the statement that there is no natural Hilbert space to describe quantum field theory in an open universe holds
generically, that is in the absence of any special assumptions about what is happening at spatial infinity.   With suitable conditions at infinity (such as asymptotic
flatness), a natural Hilbert space is sometimes available.
A typical example in which a natural Hilbert space does not exist is an open Big Bang cosmology -- for
example an expanding universe with spatial sections of zero or negative curvature.   In such a case, quantum dynamics must be described by evolution of density matrices
(or more precisely of the Type III generalization of density matrices), not by operators on a Hilbert space.  

This difference between  a closed universe and  a generic open one might come as a surprise, but  actually has little
to do with any details of quantum field theory.   The phenomenon results purely from the fact that, even with an ultraviolet cutoff in place, the number of
degrees of freedom  in an open universe is infinite.   Thus a rather similar phenomenon occurs in the thermodynamic limit of quantum statistical mechanics.
In finite volume, a quantum field theory has a Hilbert space $\H$, and thermodynamic functions such as entropy, energy density, etc., can be computed
by studying a density matrix  $\rho=\frac{1}{Z}e^{-\beta  H}$ on $\H$ (as usual, $H, \beta,$ and $Z$ are the Hamiltonian, inverse temperature, and
partition function).    Thermodynamic functions and correlation functions can be computed in finite volume and they have a large volume limit.  However, if one aims
to describe statistical mechanics directly in the large volume limit, one runs into the fact that at nonzero temperature, the Hilbert space $\H$ does not
have a large volume limit, for reasons similar to what happens in quantum field theory
in an open universe.   There is a cure for this problem  \cite{HHW}, which is to use the thermofield double, namely $\rho^{1/2}=\frac{1}{\sqrt Z}e^{-\beta H/2}$, regarded not as an operator on $\H$
but as a vector in the tensor product of two copies of $\H$.   Starting with the thermofield double state $\Psi_\TFD=\rho^{1/2}$, one can build a Hilbert space 
$\H_\TFD$ that
has an infinite volume limit, and such that thermodynamic functions and expectation values in the infinite volume thermal ensemble can be computed
as correlation functions of  operators acting in $\H_\TFD$.   A price one pays is that the operators of the original system, when interpreted
as operators on $\H_\TFD$, naturally generate a von Neumann algebra $\A_\TFD$ of Type III, not a more familiar algebra of Type I.   
The Hamiltonian operator acting on just one of the two copies in the thermofield double is {\it not} well-defined and in particular
is not part of $\A_\TFD$.  Rather, the generator of time translations acts on both copies in the thermofield double and generates an outer automorphism of $\A_\TFD$.

The need to go to the thermofield double in the thermodynamic limit of quantum statistical mechanics is quite analogous to the need to use density matrices
(or their Type III analog) 
 for quantum fields in a generic open universe.   There is also a close analogy between the reasons for the occurrence of Type III algebras.  In each of the
 two cases, the Type III nature of the algebra results from   a divergent amount of entanglement.  
In statistical mechanics, there is divergent entanglement between the two copies of the original system that make up the 
thermofield double, and in quantum field theory in curved spacetime,
there is divergent entanglement between modes inside and outside a given bounded region of space.   One divergence is an infrared divergence and one is an ultraviolet
divergence, but nonetheless they are quite similar.    Indeed  in the case of a conformal field theory, one can make a direct conformal 
mapping between the two types of divergence.

In the course of our study of the thermofield double, we will become familiar with the basic facts about (hyperfinite) von Neumann algebras, which come in three types,
Type I, Type II, and Type III. All of these algebras can be constructed from simple qubit systems.  The familiar algebras are of Type I.   As already mentioned, 
the local algebras of quantum field theory, and likewise  the algebras of observables in the infinite volume limit of quantum statistical mechanics, are of Type III.  

 In section \ref{QFT}, we provide a gentle introduction to the facts about quantum field theory in curved spacetime that were summarized earlier.
 The aim is to motivate the statements and help the reader gain an intuition.   In section \ref{Thermo}, we discuss in a similar spirit the 
 thermodynamic limit of quantum statistical  mechanics and the thermofield double.
This article was originally planned as an exposition of those matters, along the lines of 
a lecture at the 2021 Bootstrap Summer School, where the most important points are outlined  \cite{Lecture}.
  However, a recent article by  Leutheusser and Liu \cite{LL}, with some novel observations about quantum black holes,  suggested
that it would be worth while to discuss  from a similar point of view the large $N$ limit of gauge theory.  These issues are briefly discussed
in section \ref{large}.   More detail on some aspects, including a role for algebras of Type II in the presence of gravity,  will appear elsewhere \cite{WittenNew}.

In analyzing  quantum field theory in Lorentz signature, we will not make use of knowledge about what happens in Euclidean signature.   This
is in common with the existing literature, and there is a simple reason for it:  a generic Lorentz signature spacetime does not have any useful Euclidean continuation.
However, Euclidean field theory is powerful, and it is a pity not to  be able to apply this power to the Lorentz signature case.   A recent analysis of
quantum field theory on a spacetime with a complex metric \cite{KS} offers hope that it will be possible to bridge the gap and apply Euclidean-style reasoning
to the Lorentz signature case.
 
 \section{Quantum Field Theory in Curved Spacetime}\label{QFT}
 
 \subsection{The Problem}\label{problem}
 
 As a motivating example of the problem of quantum field theory in curved spacetime, 
 consider a free field theory coupled to a background metric $g$ of Lorentz signature $-++\cdots+$ on a $D$-manifold $M$.
 For example, a real scalar of mass $m$ is defined by an action 
 \be\label{scaction} I=-\frac{1}{2}\int\d^D x \sqrt g \left(g^{\mu\nu}\partial_\mu\phi\partial_\nu\phi +m^2\phi^2\right). \ee
 
 We assume that $M$ is a globally hyperbolic manifold with complete Cauchy hypersurface $S$.   The field $\phi$ obeys a second order wave equation
 \be\label{secondorder}(D^\mu D_\mu+m^2)\phi(\vec x,t)=0,\ee
 and initial data consist of the field $\phi$ and its normal derivative $\dot\phi$ along $S$.   These initial data are supposed to satisfy the usual canonical
 commutation relations $[\dot\phi(x),\phi(y)]=-\i \delta(x,y)$, with other commutators vanishing.    
 
 One of the most important facts about quantum mechanics is that given 
 any {\it finite} set of canonical variables $p_i,x^j, ~i,j=1,\cdots,d$ with the usual commutation relations
 $[p_i,x^j]=-\i \delta_i^j$, $[p_i,p_j]=[x^i,x^j]=0$, the Hilbert space $\H$ that provides an irreducible representation of this algebra exists and is unique up to isomorphism.
 $\H$ can be realized as a space of square-integrable functions (more canonically, half-densities) of $\vec x$, or as a space of square-integrable functions of $\vec p$.
 One can also work with holomorphic functions of $z_i=x^i+\i p_i$.  Many slight generalizations and combinations of these constructions are also possible.
These constructions can all be shown to be equivalent by Fourier transformations and more general unitary transformations. More generally, 
according to a theorem of  von Neumann \cite{vNe}, the quantization of $\R^{2d}$  is unique, up to isomorphism, as long as one requires 
that $\vec x$ and $\vec p$ satisfy the usual commutation relations.\footnote{It is essential here that $\R^{2d}$ is not considered as an abstract symplectic manifold,
but rather we are given a preferred set of linear functions $\vec x,$ $\vec p$ on $\R^{2d}$  (unique up to linear symplectic transformations and additive constants) whose commutators are specified.
The problem of quantizing $\R^{2d}$ as an abstract symplectic manifold with no additional structure is completely different, and has no natural solution, because
of the operator ordering problem of quantum mechanics.   In our field theory problem, the phase space comes with a distinguished set of linear functions,
namely the field variables $\phi(x)$ and $\dot\phi(x)$.}
All of these constructions produce a distinguished Hilbert space $\H$.   There is no distinguished 
vector in $\H$ unless more structure is present, for example, a Hamiltonian whose ground state would be such a distinguished vector.

In infinite dimensions,  this uniqueness of the irreducible representation of the canonical commutators is completely false \cite{GW}.  We will see that presently
 with  examples.   Though we primarily discuss bosons in this article, the same statements hold for fermions:   the canonical anticommutators have an
 essentially unique representation in the case of finitely many fermionic variables, but not in the case of an infinite set of fermionic variables.
 
Once one picks a  representation of the field variables on  $S$, satisfying the canonical commutation relations, 
 in a Hilbert space $\H$, the rest of the construction of the theory is straightforward.   
The field $\phi$ obeys the second order wave equation $(D_\mu D^\mu+m^2)\phi=0$, by virtue of which $\phi(\vec x,t)$ can be expressed for
all $\vec x,t$  in terms of $\phi,\dot\phi$
on $S$.   Such an expression exhibits $\phi(\vec x,t)$ as an operator acting on $\H$, and therefore matrix elements $\la\Psi|\phi(\vec x_1,t_1)\phi(\vec x_2,t_2)
\cdots \phi(\vec x_n,t_n)|\chi\ra$, for states $\Psi,\chi\in\H$, are uniquely determined, in principle.     However, if one starts this construction with the ``wrong''
representation of the canonical commutation relations, then the resulting matrix elements $\la\Psi|\phi(\vec x_1,t_1)\phi(\vec x_2,t_2)
\cdots \phi(\vec x_n,t_n)|\chi\ra$ are not physically sensible: they do not have the expected short distance singularities.   From this point of view, the problem
of quantization is to find the right representation of the canonical commutators such that the resulting correlation functions are physically sensible.

 One simple case in which it is obvious on physical grounds how to construct the appropriate representation of the canonical commutation relations
 is a spacetime with a Killing vector field that is everywhere timelike.   In a suitable coordinate system $t, \vec x$, the metric of such a spacetime
 is independent of the time $t$.     In this case, we expect the quantum theory to have a self-adjoint Hamiltonian operator $H$
that generates time translations. In particular, such an $H$ is diagonalizable.  Because the Killing vector field is everywhere timelike, we expect $H$ to be bounded below.
A representation of the canonical commutation relations that admits such an $H$ is  unique up to isomorphism.
   To construct this representation, 
  we simply expand the field in $c$-number modes $f_k(\vec x)$, $\bar f_k(\vec x)$ 
 of positive and negative frequency\footnote{The sum here is a discrete sum in a closed universe, and becomes a continuous
 integral in an open universe.  We assume for the moment that all modes of $\phi$ have positive or negative frequency; zero-modes will be incorporated presently.}
 \be\label{expfields}\phi(\vec x,t)=\sum_k \left( a_k \,f_k(\vec x)
 e^{-\i\omega_k t} +a^\dagger_k \bar f_k(\vec x) e^{\i\omega_k t}\right),~~~\omega_k>0.\ee
 The modes can be normalized so that the operators $a_k$, $a_k^\dagger$ obey the canonical
 relations $[a_k,a_l]=[a^\dagger_k,a^\dagger_l]=0$, $[a_k,a^\dagger_l]=\delta_{kl}$.  
 Acting with $a^\dagger_k$ increases the energy (the eigenvalue of $H$) by $\omega_k$, and acting with $a_k$ reduces the energy
 by the same amount.
  We introduce a ``ground state'' $\Omega$
 of lowest energy, annihilated by the $a_k$.      Then we define a pre-Hilbert space $\H_0$ of all states that can be created by acting on 
 $\Omega$ with finitely many $a^\dagger$'s.   To be more precise, we define $\H_0$ to have a basis of states that we denote as
 \be\label{basiss} a_{k_1}^\dagger a_{k_2}^\dagger \cdots a_{k_n}^\dagger|\Omega\rangle,~~~n=0,1,2\cdots.\ee
 The $a$'s and $a^\dagger$'s are defined to act on these states in the familiar way, satisfying the commutation relations.   
 $\H_0$ has all the properties of a Hilbert space except
 completeness; that is why we call it a pre-Hilbert space.   We take the Hilbert space closure of $\H_0$ and this gives us the
 desired Hilbert space $\H$ for the free field $\phi$ in a time-independent curved spacetime.  Clearly, $\H$ contains a distinguished vector of minimum energy,
 namely $\Omega$, which is annihilated by all the $a_k$.  Conversely, any Hilbert space $\H$ that represents the canonical commutators and in which $H$ is
 self-adjoint and bounded below must contain a vector $\Omega$ annihilated by all the $a_k$ (otherwise acting repeatedly with the $a_k$ on an eigenvector
 of $H$ would lower the energy indefinitely), and therefore is equivalent to the Hilbert space that was just constructed.

 This discussion must be slightly modified if there are zero-modes.    For a simple example with zero-modes, simply set $m=0$ in the model we have been discussing.   Then the field $\phi$ does have a zero-mode, namely the constant mode $\phi_0$.  The canonical conjugate to $\phi_0$
 is the constant mode $\pi_0$ of $\dot\phi$.   If $\pi_0$ is suitably normalized, the commutation relations of these modes are the usual canonical commutators
 $[\pi_0,\phi_0]=-\i$.  This is a finite set of canonical variables, so the Hilbert space $\H'$ obtained by quantizing
 these modes is unique, up to isomorphism.  It can be defined, for example, as the space of $L^2$ functions of $\phi_0$, with $\pi_0$ acting by
 $-\i\partial/\partial\phi_0$.     The unique representation of the canonical commutators of the full system in which $H$ is self-adjoint and bounded below is then given by the
 Hilbert space $\H\otimes\H'$, where $\H$ is obtained by quantizing the nonzero modes as in the last paragraph and $\H'$ is obtained by quantizing the zero-modes.
 More generally, in a time-independent closed universe, the number of zero-modes, in a free field theory with fields of any spin, is always finite,\footnote{For example,
 in $U(1)$ gauge theory with gauge field $A$, zero-modes correspond to harmonic 1-forms on $S$, and the number of such modes is the first Betti number of $S$,
 which is always finite for compact $S$. In this example, the Hilbert space has different components labeled by the first Chern class of the line bundle
 on which $A$ is a connection.   Each component can be analyzed as we have done in the text for the massless scalar.}   so the possible existence
 of zero-modes never  presents any problem in quantization.
 
 Note, however, that in the example with the massless scalar field, the Hilbert space $\H'$ that is obtained by quantizing the zero-modes does not have any distinguished
 vector.   The Hamiltonian $H$ acts in $\H'$ as a multiple of $\pi_0^2=-\partial^2/\partial\phi_0^2$.   This operator does not have a normalizable ground state
 that would provide a distinguished vector in $\H$.   So this is a simple example of successfully quantizing a theory in curved spacetime and not finding any
 distinguished vector in the resulting Hilbert space.   That is somewhat exceptional in a time-independent spacetime, since it only happens in the presence
 of zero-modes (and not always then).   But when one relaxes the assumption of time-independence, it is the typical state of affairs.

In a time-dependent situation, it is less obvious how to select an appropriate representation of the canonical commutation relations.
Before discussing this question, we will practice with the simpler case  of a spin system.  This will also provide useful background
when we come to statistical mechanics in section \ref{Thermo}.

 \subsection{Practicing With A Spin System}\label{spins}
 
 A qubit is a two-state quantum system, comprising, possibly, the two states of a spin 1/2 particle.  A single qubit realizes the algebra of Pauli matrices
 $\sigma_x, \sigma_y, \sigma_z$. So a system of $n$ qubits realizes the corresponding algebra of operators $\sigma_{a,k}$ where $a=x,y$ or $z$ and $k\in\{1,\cdots,n\}$
 labels the choice of qubit:
 \be\label{toffox}\sigma_{a,k}\sigma_{b,k}=\delta_{ab}+\i\epsilon_{abc}\sigma_{c,k},~~~~ [\sigma_{a,k},\sigma_{b,k'}]=0,~~k\not=k'.\ee
 For finite $n$, this algebra has, up to isomorphism, a unique irreducible representation, of dimension $2^n$.   But for an infinite set of qubits, the representation
 of the algebra becomes highly non-unique.  That is the point that we will explore here.\footnote{The following matters are discussed, for example,
 in section 2.3 of \cite{Sewell2}. That book is also a useful reference for other topics relevant to the present article.}

As a preliminary, consider this  question.   Consider a countably infinite set of qubits (with no additional structure).
Does the Hilbert space of
 such a system have a countable or uncountable dimension?
 
 A Hilbert space $\H^*$ that literally describes {\it all} states of a countably infinite collection of qubits definitely has an uncountable
 dimension.   But it is normally not useful to do physics in a Hilbert space as big as that.   Infinite
 constructions in physics are normally limits of finite constructions, and it is generally  possible to take the limit in such a way
 that all questions of interest can be answered in a Hilbert space of countably infinite dimension.   Such a Hilbert space is said to be ``separable.''
 What we should do to get a separable subspace of $\H^*$ depends on what we are interested in.   To get anywhere,
 we should have a class of operators on $\H^*$ that we regard as sensible physical observables.  Then we can look
 for a separable subspace $\H$ of the ridiculously big Hilbert space $\H^*$ that is big enough to realize the algebra of
 physical observables and to describe the states that we care about.  There are two ingredients here: what class of observables
 do we care about and what class of states do we care about?   The two cannot be specified independently since the observables
 we care about have to be able to act on the states that we care about.

 In the case of the qubits, we might decide that we want a class of observables that contains
 all the single qubit operators.   A basis of operators on the $k^{th}$ qubit (apart from the
 identity) are the Pauli spin operators $\sigma_{a,k}$.  We define an algebra $\A_0$ that consists of all polynomials in these operators.
 Thus $\A_0$ consists of finite linear combinations of products of the form
 $\sigma_{a_1,k_1}\sigma_{a_2 ,k_2}\cdots \sigma_{a_r,k_r}$, satisfying the relations (\ref{toffox}).    The reason to call this $\A_0$ is that
 usually one wants some kind of completion of $\A_0$ to get the algebra $\A$ of physical
 observables.  But we cannot discuss the completion without deciding what kind of states we are
 interested in.
 
 For example, we might be interested in states in which the qubits almost all have ``spin up'' along the
 $z$-axis, or in other words in which they are almost all in the state $\left|\uparrow\right\rangle= \left| \begin{matrix} 1\cr 0\end{matrix}\right\rangle$. 
 Then we might decide that the physical Hilbert space $\H$ should at least contain the vector
 \be\label{thevector}\Omega_\uparrow=\otimes_n \left|\uparrow\right\rangle_n. \ee
 If we want to have $\A_0$ as an algebra of observables, we have to include all states obtained by
 acting with $\A_0$ on $\Omega_\uparrow$.   When we act with $\A_0$ on $\Omega_\uparrow$,
 since $\A_0$ consists of all the operators that act on a finite set of qubits, we get all states in which all
 but finitely many qubits are in the state $\left|\uparrow\right\rangle$.  States of this kind make
 a pre-Hilbert space $\H_0$ with a countable basis.   Its Hilbert space completion $\H$ is separable.
 Moreover, we can now actually take a completion of $\A_0$ by adding convergent limits of sequences
 of operators in $\A_0$.  (The relevant notion of convergence is discussed momentarily.)
  This will give the algebra $\A$ of all bounded operators on $\H$.    
 By definition, the algebra of all bounded operators on a Hilbert space $\H$ is called a von Neumann algebra
 of Type I.   Von Neumann algebras of Type I
 are the familiar ones, but we will meet the other types. 
 
Obviously, instead of the state $\Omega_\uparrow$, we could have started with another state in the very
large Hilbert space $\H^*$.  For example, setting $\left|\downarrow\right\rangle= \left| \begin{matrix} 0\cr 1\end{matrix}\right\rangle$,
we could have started with the state $\Omega_\downarrow = \otimes_n \left|\downarrow\right\rangle_n$ in which the spins
are all down.   Then we would define a pre-Hilbert space $\H_0'$ in which all but finitely many spins are down; it has a  Hilbert space
completion $\H'$.   The algebra $\A_0$ acts on $\H'$, and by including convergent limits of operators,
it  can again be completed to a von Neumann algebra $\A'$ that in this example will be the Type I algebra of all bounded
operators on $\H'$.

The algebras $\A$ and $\A'$ are actually different, but to understand why we need to explain what we mean by saying that
a sequence $\a_1,\a_2,\cdots\in \A_0$ converges.   If $\A_0$ is understood as an algebra of bounded operators on a Hilbert space $\H$,
the relevant notion of convergence (which mathematically is called a weak limit) is to say that the sequence $\a_1,\a_2,\cdots \in \A_0$ converges
if the limit
 $\lim_{n\to\infty}\langle\psi|\a_n|\chi\rangle$ exists, for all $\psi,\chi\in\H$.   In that case, one defines
 an operator $\a$ whose matrix elements are $\langle\psi|\a|\chi\rangle=\lim_{n\to\infty}\langle\psi|\a_n|\chi\rangle$,
 and one says that $\lim_{n\to\infty}\a_n=\a$. 
 The explanation given by Haag \cite{HaagBook} for why this is the appropriate notion of convergence for physics
  is that any given experiment measures finitely many matrix elements to finite precision, and therefore,
if $\a$ is the weak limit of a sequence $\a_1,\a_2,\cdots,$ then any given experiment cannot distinguish $\a_n$ from $\a$ if $n$ is sufficiently
 large.   The definition of a weak limit, however, makes it clear that to determine the limit of a sequence of elements in $\A_0$, we need to know
 the Hilbert space $\H$ on which $\A_0$ is acting.   For a concrete example, let $\sigma_{+,n}=\frac{1}{2}\left(\sigma_{x,n}+\i \sigma_{y,n}\right)$ be the spin raising operator
 for the $n^{th}$ qubit.    Then $\lim_{n\to\infty}\sigma_{+,n}^\dagger\sigma_{+,n}=0$  if the operators are considered to act on $\H$, but
 $\lim_{n\to\infty}\sigma_{+,n}^\dagger\sigma_{+,n}=1$ if the operators are consisted to act on $\H'$.  These statements reflect the fact for any state in $\H$, the
 $n^{th}$ spin is almost certainly up (and annihilated by $\sigma_{+,n}$) for sufficiently large $n$, while for any state in $\H'$, it is almost certainly down.     As a matter of terminology, an algebra of operators that is closed under weak limits is called a von Neumann algebra.
 
 Instead of considering an abstract collection of qubits with no structure, we could arrange the qubits on a lattice of some dimension
 and consider a lattice Hamiltonian $H$, for instance a Heisenberg spin Hamiltonian.   Then it would be natural
  to take $\Psi$ to be the ground state of this Hamiltonian, assuming it is unique.  Starting with this
 $\Psi$, we would then construct a separable Hilbert space $\H$ on which all of the elementary spin operators
 act.   In this Hilbert space, we would complete the algebra $\A_0$ to a von Neumann algebra $\A$ acting on $\H$. 
 This algebra will be of Type I.  Our previous discussion amounted
 to the special case that $H=-\sum_n \sigma_{z,n}$, leading to the ground state
 $\Psi=\Omega_{\uparrow}$, or $H=+\sum_n\sigma_{z,n}$,
 leading to $\Psi=\Omega_{\downarrow}$.
  If $H$ has degenerate ground states,
 then one would want to pick a basis of ground  states consisting of states that satisfy cluster decomposition, and proceed as before,
 starting with any one of those states.  In a moment, we will explain what is special about the states that satisfy cluster decomposition.
 
We can similarly construct a separable Hilbert space with an action of a completion of the algebra $\A_0$ starting with any vector  $\Psi\in\H^*$.
There is a certain sense, however, in which it is not true that every $\Psi$ is an equally good starting point.   To see this,
consider the vector $\Psi=\frac{1}{\sqrt 2}\Omega_\uparrow +\frac{1}{\sqrt 2}\Omega_\downarrow.$
Let $\pi_\uparrow$ and $\pi_\downarrow$ be the operators that project the first qubit onto the states $|\negthinspace\uparrow\ra$ and $|\negthinspace\downarrow\ra$,
respectively.   Then $\pi_\uparrow \Psi=\frac{1}{\sqrt 2}\Omega_\uparrow$ and $\pi_\downarrow\Psi = \frac{1}{\sqrt 2}\Omega_\downarrow$.
Further acting with arbitrary elements of $\A_0$, we see that any state in either $\H_\uparrow$ or $\H_\downarrow$ can be approximated
by states  $\a\Psi$, $\a\in\A_0$,
and therefore $\H_\Psi=\H_\uparrow\oplus \H_\downarrow$.   Thus, while  it is true that 
by acting with $\A_0$  on the state $\Psi$, we can build a separable Hilbert
space $\H_\Psi$ that admits an action of a completion $\A$ of  $\A_0$, we see that $\A$ does not act irreducibly on $\H_\Psi$,
since $\H_\Psi$ has the $\A$-invariant decomposition $\H_\uparrow\oplus \H_\downarrow$.     If $\Psi$ is such that a completion of
$\A_0$ does act irreducibly on $\H_\Psi$, we say that $\Psi$ is a pure state\footnote{There is some tension here
with the usual terminology in quantum mechanics, where the phrase ``pure state'' usually refers to any Hilbert space state,
without such a restriction.  The point is that if $\A$ is a Type I algebra, consisting of all bounded operators on a Hilbert space $\H$,
then every $\Psi\in\H$ is a pure state in the algebraic sense (the action of $\A$ on $\Psi$ generates the whole Hilbert space $\H$,
on which $\A$ acts irreducibly).   Thus for the case that $\A$ is of Type I, the two notions of ``pure state'' are compatible.}
 for the algebra $\A_0$.   The pure states are the most interesting ones since they correspond to the irreducible representations of the 
 relevant algebras of observables.   In the example of a lattice spin Hamiltonian with multiple 
ground states, the pure states are the ones that satisfy cluster decomposition.

This description of what is meant by a pure state for the algebra $\A_0$ is not the most common one.   Usually, one considers the function 
$F_\Psi(\a)=\la\Psi|\a|\Psi\ra$, which obeys the following properties: 
\begin{itemize}\item{  it is linear in $\a$, namely $F(\lambda \a+\mu \b)=\lambda F(\a)+\mu F(\b)$, for $\a,\b\in\A_0$, $\lambda,\mu\in \C$,} \item{ it is normalized
to $F(1)=1$,} \item{  it is positive, in the sense that  $F(\a^\dagger \a)\geq 0$ for all $\a\in\A$.} \end{itemize}
A function on an algebra, in the present example $\A_0$, that satisfies these properties is called a state on the algebra.      Clearly then for any $\Psi\in\H^*$, the function
$F_\Psi(\a)=\la\Psi|\a|\Psi\ra$ is a state on the algebra $\A_0$.   A  
state on an algebra is called ``pure'' if it is not a convex linear combination of
other states, or in other words if it is not possible to write $F(\a)= p_1 F_1(\a)+p_2 F_2(\a)$ with $p_1,p_2>0$ and states $F_1$, $F_2$.
In the present example,  one readily sees that
\be\label{knicky} F_\Psi(\a)=\frac{1}{2} \la\Omega_\uparrow|\a|\Omega_\uparrow\ra+\frac{1}{2} \la\Omega_\downarrow|\a|\Omega_\downarrow\ra,
~~\a\in \A_0.\ee
So the state $F_\Psi$ is not pure.   The two notions of a pure state that we have described are 
equivalent: $\H_\Psi$ is irreducible for a completion of $\A_0$ if and only if the state $F_\Psi(\a)$ is pure.   The interested reader
can verify this, possibly after reading about the GNS construction in section \ref{td}.

 Suppose  that we construct a separable Hilbert space $\H_\Psi$  as before from some
 input state $\Psi\in\H^*$, and another Hilbert space $\H_{\Psi'}$ starting with a different input state $\Psi'$.
 $\H_\Psi$ and $\H_{\Psi'}$ are both subspaces of the very big Hilbert space $\H^*$.   Are they the same
 subspaces?   A necessary condition is clear: $\Psi'$ should be in $\H_\Psi$.   If so, then  $\Psi'$
 can be approximated by states  $\a\Psi$, with $\a\in \A_0$.    But in that case, any vector in $\H_{\Psi'}$,
 which by definition can be approximated by $\a'\Psi'$ with $\a'\in\A_0$, can in fact be approximated by $\a'\a\Psi$ and hence is
 in $\H_\Psi$.   In other words, if $\Psi'\subset\H_\Psi$ then $\H_{\Psi'}\subset \H_\Psi$.     If $\Psi$ is pure, it actually
 follows that $\H_{\Psi'}=\H_\Psi$.   That is because if $\H_{\Psi'}\subset\H_\Psi$, then $\H_\Psi$ 
  has an $\A_0$-invariant decomposition $\H_\Psi=\H_{\Psi'}\oplus \H_{\Psi'}^\perp$.   If $\Psi$ is a pure state
 for the algebra $\A_0$, then one of the summands here must be trivial and so in this case $\H_\Psi=\H_{\Psi'}$.

 \subsection{A System Of Harmonic Oscillators}\label{harmo}
 
 Now let us discuss something that is a little closer to field theory.    We will consider an infinite sequence
 of pairs of canonical variables, $(x_1,p_1),$ $(x_2,p_2)$, $\cdots$.    The $k^{th}$ pair can be quantized to give
 a Hilbert space $\H_k$, and for this, there is no need to know anything about a preferred vector.   If we want to describe
 all of the states in all of the $\H_k$, we have to define a Hilbert space
 \be\label{bigone}\H^*=\otimes_k \H_k\ee
 of uncountably infinite dimension.   But if we are given a state $\Psi$ that for some reason we like, we can proceed
 as we did with the qubits.   We define an algebra $\A_0$ consisting of all polynomials in the $x$'s and $p$'s.  Then we define
 a pre-Hilbert space $\H_0$ consisting of all states that can be made from $\Psi$ by acting with $\A_0$,
 and as usual we complete this to a Hilbert space $\H$.  We can then also complete $\A_0$ to an algebra $\A$.
 
 A simple example of this would be that $\Psi$ is a tensor product of states $\psi_k\in \H_k$:
 \be\label{welfo}\Psi=\otimes_k\psi_k. \ee
 We can assume that the $\psi_k$ and therefore also $\Psi$ are normalized, $(\psi_k,\psi_k)=(\Psi,\Psi)=1$.
 We construct a Hilbert space $\H_\Psi$ by acting with $\A_0$ on  $\Psi$.   On the other hand, consider another
 state of the same kind, $\Psi'=\otimes_k\psi'_k$, where again we assume the $\psi'_k$ and $\Psi'$ to be normalized.
 Is the Hilbert space $\H_{\Psi'}$ the same as $\H_\Psi$?  
 
 Let $c_k=(\psi'_k,\psi_k)$.  After possibly changing the phases of the $\psi'_k$, we can assume that $c_k\geq 0$;
 clearly $c_k\leq 1$.   We have
 \be\label{dotto} |\langle\Psi',\Psi\rangle|=\prod_{k=1}^\infty c_k. \ee
 We consider separately two cases: (1) $\langle\Psi',\Psi\rangle\not=0$; (2) $\langle\Psi',\Psi\rangle=0$.
 In case (1), the $c_k$ rapidly converge to 1.   By acting with the algebra $\A_0$ on $\Psi$, we can
 change any finite set of $c_k$'s at will, subject to the condition $c_k\leq 1$, without changing the normalization condition
 $(\psi_k,\psi_k)=1$.  In particular, we can find a sequence $\a_1,\a_2,\cdots\in\A_0$ such that the states $\a_k\Psi$ are
 all normalized and
 $\lim_{n\to\infty}\la\Psi',\a_k\Psi\ra=1$.   In other words, $\Psi'$ can be approximated by vectors $\a_k\Psi$; hence $\Psi'\in\H_\Psi$.
 The relation between $\Psi$ and $\Psi'$ was symmetrical, so likewise $\Psi\in\H_{\Psi'}$.
 As in section \ref{spins}, it follows that $\H_\Psi=\H_{\Psi'}$.
 
 In case (2), where $\prod_{k=1}^\infty c_k$ vanishes, the interesting case is that the $c_k$ are not individually 0 and
 $\prod_{k=1}^\infty c_k=0$ because the $c_k$ do not approach 1 fast enough to make this product nonzero.   
 Then this vanishing cannot be changed by changing finitely many of the $c_k$.   So in such a case, $\Psi'$ is
 orthogonal not just to $\Psi$ but to $\a\Psi$ for all $\a\in \A_0$.   It follows then that $\a'\Psi'$ is orthogonal to $\a\Psi$
 for all $\a,\a'\in\A_0$, since $\la \a'\Psi'|\a\Psi\ra=\la\Psi'|\a'{}^\dagger \a\Psi\ra=0$.   So $\H_{\Psi'}$ and $\H_\Psi$ 
 are orthogonal.   
 
If finitely many of the $c_k$ vanish, those $c_k$'s should just be disregarded since they could be set to 1 by
replacing $\Psi$ with some $\t\Psi=\a\Psi$, $\a\in\A_0$, without changing the other $c_k$'s.    So in case
finitely many of the $c_k$ vanish, we look at the restricted product $\prod'_k c_k$ over the nonzero $c_k$;
we will get $\H_{\Psi'}=\H_\Psi$ if this restricted product is positive, and otherwise $\H_{\Psi'}$ is orthogonal to $\H_\Psi$.
If infinitely many $c_k$ vanish, we cannot change this by acting with an element of $\A_0$, so $\H_{\Psi'}$ and $\H_\Psi$
are orthogonal.

It is noteworthy that there are only two outcomes here: $\H_\Psi$ and $\H_{\Psi'}$ are the same, or they are orthogonal.\footnote{We did not
run into other possibilities encountered in section \ref{spins}, because the states $\Psi$ and $\Psi'$ are pure for the algebra $\A_0$. This resulted from their product structure.}
For $\H_\Psi$ and $\H_{\Psi'}$ to be the same, we require no relation at all between $\psi_k$ and $\psi'_k$ for all but finitely
many modes, but for large $k$, $\psi_k$ and $\psi'_k$ must coincide asymptotically at a sufficiently fast rate.

Now we will consider a special case of this that is useful background for field theory.   For each $k$, pick a complex number $\tau_k$
in the upper half plane, and define
\be\label{bicy} a_k^\dagger=\frac{1}{2\sqrt{\Im\,\tau_k}}\left(x_k+\tau_k p_k\right),~~ a_k=\frac{1}{\sqrt{2\Im\,\tau_k}}\left(x_k+\bar\tau_k p_k\right). \ee
This has been chosen to ensure that $[a_k^\dagger, a_{k'}]=\delta_{kk'}$, with other commutators vanishing.    We can  choose
$\psi_k$ so that $a_k \psi_k=0$, whereupon if $\Psi=\otimes_k\psi_k$, we have $a_k\Psi=0$ for all $k$.   We can view $\Psi$
as the ground state of a harmonic oscillator Hamiltonian $H=\sum_k \epsilon_k a_k^\dagger a_k$, for arbitrary $\epsilon_k>0$.

On the other hand, we could make the same construction with a different set of parameters $\tau'_k$, leading to a different set of creation
and annihilation operators $a'_k{}^\dagger,$ $a'_k$ and a different vector $\Psi'=\otimes_k \psi'_k$ that is annihilated by the new
annihilation operations.  Now we can ask if the Hilbert spaces $\H_\Psi$ and $\H_{\Psi'}$ are the same.   From the above analysis,
the condition for this is $\prod_{k=1}^\infty c_k>0$, with $c_k=\la\psi'_k,\psi_k\ra$.   This is equivalent to a condition that $\tau'_k$ is sufficiently close to $\tau_k$
for large $k$, with no condition at all on $\tau'_k$ for any finite set of values of $k$.   

For any given $k$, at the classical level, a group $SL(2,\R)\cong Sp(2,\R)$ of linear canonical transformations acts on the pair $x_k,p_k$,
preserving the commutation relations. The group that acts on the quantum Hilbert space is a double cover of $Sp(2,\R)$,
called the metaplectic double cover.\footnote{One way to see the occurrence of this double cover is the following.   For a pair of canonical
variables $x,p$, let $H$ be the harmonic oscillator Hamiltonian $H=\frac{1}{2}(p^2+x^2)$.  As a canonical transformation of $x$ and $p$, $\exp(2\pi\i H)=1$.
But quantum mechanically, as the eigenvalues of $H$ are all half-integers, one has $\exp(2\pi\i H)=-1$.} We denote the double cover
as $\widehat{Sp}(2,\R)$.  It is possible to choose for each $k$ an element
$g_k\in \widehat{Sp}(2,\R)$ that conjugates $a_k^\dagger$ and $a_k$ to $a'{}^\dagger_k$ and $a'_k$, and maps $\psi_k$ to $\psi'_k$.   
Clearly, any finite product of the $g_k$'s acts on the Hilbert space $\H_\Psi$ (and likewise on $\H_{\Psi'}$).   But the infinite
tensor product $\otimes_{k=1}^\infty g_k$ maps $\Psi$ to $\Psi'$, so it maps $\H_\Psi$ to $\H_{\Psi'}$.   It only acts within $\H_\Psi$
if $\prod_{k=1}^\infty c_k>0$, which is equivalent to saying that $g_k$ approaches 1 sufficiently rapidly for $k\to\infty$.

The moral of the story is that in the case of an infinite set of canonical variables, a linear canonical transformation is realized in
a given representation of the canonical commutation relations if and only if it is sufficiently close to the identity on all but finitely
many of the  variables.   To motivate this statement, we have considered the simple case of a canonical transformation that,
in a suitable basis, acts separately on each pair of variables $x_k,p_k$.   However, the conclusion is general and is explained in generality,
 for example, in \cite{WaldBook}.

Another way to state the conclusion is that for an infinite set of canonical variables, to determine a specific representation of the canonical
commutators in a separable Hilbert space $\H$, one needs to  describe the desired representation to sufficient precision for all but finitely many variables.  This may be done by decomposing the field variables in creation and annihilation operators, where $\H$ is supposed
to contain a vector annihilated by the annihilation operators.  To specify a particular $\H$ in this way, one must specify the decomposition
in creation and annihilation operators in an asymptotically precise way.

 \subsection{Back To Field Theory}\label{fff}
 
 Now we go back to quantum field theory in curved spacetime.   First we consider a closed universe, that is a globally
 hyperbolic spacetime $M$ with compact Cauchy hypersurface $S$.     Modes along $S$
 of very short wavelength compared to the radius of curvature of
 $S$  can be separated to positive and negative
 frequency, with an uncertainty that rapidly vanishes in the limit of short wavelengths.  
So for very short wavelength modes, there is a natural decomposition in creation and annihilation operators to good approximation, and this approximation becomes asymptotically precise
in the limit of short wavelengths.    For low energy modes, we have no notion
 at all of how to make a decomposition in creation and annihilation operators, or how to pick a preferred state in any other way.   But in a closed
 universe, there are only finitely many low energy modes.   
 
Therefore, we are in a good situation.   For all but finitely many low energy modes, we have a good approximate notion of how to make
a separation in creation and annihilation operators.  Moreover this notion is asymptotically precise for the infinitely many modes of asymptotically
short wavelength.  Such an asymptotic separation in creation and annihilation operators is precisely sufficient to define a distinguished Hilbert
space $\H$ with an action of the canonical commutation relations.  The Hilbert space $\H$ constructed this way does not really depend on the
choice of $S$, because at asymptotically small wavelengths, evolution of the field $\phi$ via its equations 
of motion preserves the separation in positive and negative frequencies. So the asymptotic separation in positive and negative
frequencies made on one surface $S$ agrees with the asymptotic separation that one would make on another surface $S'$.
 Moreover,  matrix elements of products of field operators
between vectors in $\H$ have the standard short distance singularities, because $\H$ was defined using the standard separation in 
positive and negative frequencies at short distances.   

A  rigorous version of this  argument is described on pp. 96-7 of \cite{WaldBook}, roughly as follows.   Without changing the metric
of $M$ near $S$, we can modify it to be time-independent sufficiently far to the past of $M$ and sufficiently far to the future of $M$.
Let $g_-$ be the time-independent metric in the past and $g_+$ the time-independent metric in the future.
 As explained in section \ref{problem}, using the time-independent metric $g_-$ in the past, we can straightforwardly
 construct a representation of the canonical commutation relations based on a decomposition in positive and negative frequencies.  We call the
 resulting Hilbert  space $\H_{g_-}$.  Following the same procedure in the future, we construct  a representation of the canonical commutation relations, and a Hilbert space that we  call $\H_{g_+}$.   Either choice, together with the equations of motion, determines
a representation of the canonical commutators throughout all of $M$.   
However, the two representations and the two Hilbert spaces $\H_{g_-}$ and $\H_{g_+}$ are canonically the same.
This is so because in a spacetime that is time-independent in both the past and the future, at asymptotically high energies, 
propagation from the past to the future preserves the separation in positive and negative frequencies.
The error in this statement is exponentially small at high energies, so the criterion discussed in section \ref{harmo} for two representations of
the canonical commutation relations to be the same is satisfied.  
Saying that $\H_{g_-}$ and $\H_{g_+}$ are canonically the same means that one can define transition amplitudes from a state in $\H_{g_-}$ in the past
to a state in $\H_{g_+}$ in the future, and it also means that if we restrict to a neighborhood of $S$, we get the same representation of the canonical
commutation relations and the same Hilbert space whether we use $\H_{g_-}$ or $\H_{g_+}$.  
 But as $g_-$ and $g_+$  can be varied independently, the fact that the representation of the commutation relations that we get in a neighborhood of 
 $S$ can be determined
 just from $g_-$ and can also be determined just from $g_+$ means that actually this representation does not depend on $g_-$ or $g_+$ at all.
 In other words, this construction produces a distinguished representation of the canonical commutators on $S$.
     Now we return to the original
metric on $M$, whatever it was, with no assumption of time-independence anywhere, and we use the representation of the
canonical commutation relations in a neighborhood of $S$ that was just determined.     As explained in section \ref{problem}, a choice of a representation of the
canonical commutation relations near $S$, when combined with the equations of motion, is sufficient to completely determine the theory.   This concludes the argument.

This discussion may make it clear why (as also discussed in \cite{WaldBook}) 
there is not a natural Hilbert space for a quantum field in a generic open universe, for example in a Big Bang cosmology.
In a generic open universe, there are infinitely many modes of  moderate wavelength for which we have no candidate for a preferred state
or for a separation in creation and annihilation operators.
Hence there is no natural construction of a Hilbert space.   
To construct a Hilbert space, either 
by picking  a separation in creation and annihilation operators for all the modes of moderate wavelength,
or by otherwise specifying a  seed vector $\Psi$ that could be the starting point
of the construction,  one has to supply  
a great deal of detailed information about what is happening at spatial infinity.     It is possible to define a Hilbert space that
describes a quantum field in an open universe, but there are many inequivalent constructions of such a Hilbert space and there is no natural choice.

 \begin{figure}
 \begin{center}
   \includegraphics[width=3.5in]{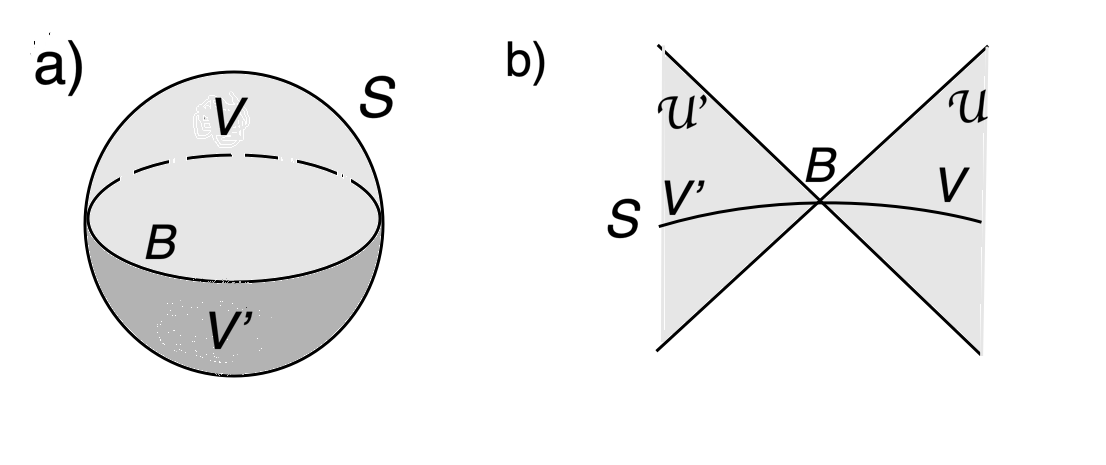}
 \end{center}
\caption{\small  (a)  A Cauchy hypersurface $S$, drawn as a two-sphere, is divided by the equator $B$ into an upper
hemisphere $V$ and a lower hemisphere $V'$.   (b)  In this view, $S$ is drawn as a one-dimensional curve
and $B$ is depicted as a point that divides $S$ into the two open sets $V$ and $V'$.   $\U$ and $\U'$ are open sets in spacetime
that are the domains of dependence of $V$ and $V'$. It is apparent from this picture that near $B$, $\U$ and $\U'$ can be modeled by opposite Rindler wedges in Minkowski space. \label{Bounded}}
\end{figure} 

\subsection{What is Quantum Field Theory In An Open Universe?}\label{whatis}

If there is no natural Hilbert space in an open universe, what can it mean to do quantum field theory in one?   
Before answering this question, let us first discuss closed universes a little more. 
Consider an observer who has access only to 
a portion of a closed universe, or who wants to develop a formalism suitable to describe experiments in just a portion of the universe.   
We will be particularly  interested in a  region of spacetime of the following sort (fig. \ref{Bounded}).   Let $B$ be a closed, codimension
1 submanifold of a Cauchy hypersurface $S$, such that the complement of $B$ is the union of two disjoint open sets $V$ and $V'$.
Let $\U$ and $\U'$  be the domains of dependence of $V$ and $V'$, respectively.   We call an open set of the form $\U$ or $\U'$
a ``local region,'' and we want to know what quantum field theory says for an observer who makes observations only in such a local region. 

Classically, in a relativistic field theory, fields in $\U$ are determined by initial data on $V$.   Quantum mechanically, this is also true
in free field theory; we exploited this fact in section \ref{problem}.  In a generic  non-free quantum field theory, it is in general not possible
to define observables on a spacelike surface such as $V$, and instead we thicken $V$ slightly to an open set $\V$, satisfying
$V\subset \V\subset \U$.   Roughly, we will describe the sense in which initial data for observations in $\U$ can be formulated in $\V$.  
Note that $\U$ and $\V$ are both globally hyperbolic spacetimes in their own right, with a Cauchy hyperurface -- namely $V$ -- which
is a portion of the Cauchy hypersurface $S$ of the full spacetime $M$.

An analog in ordinary quantum mechanics of restricting from a spacetime $M$ to a local region $\U$ is to consider only a subsystem
$X$ of a larger bipartite system $XY$.  In ordinary quantum mechanics, the subsystems $X$ and $Y$ are described by Hilbert spaces $\H_X$
and $\H_Y$, and the composite system $XY$  is described by the tensor product $\H_{XY}=\H_X\otimes\H_Y$.   
A general state of the full system $XY$, possibly a pure state, when restricted to the subsystem $X$, is described by a density matrix
$\rho_X$.   Here $\rho_X$  is a self-adjoint operator, positive and of trace 1, acting on the Hilbert space $\H_X$ of the subsystem.  
 Let $\A_X$ be the algebra of all operators of  subsystem $X$; this is the same as the algebra of all operators on $\H_X$. 
For $\a\in \A_X$,  its expectation value in a state characterized by the density matrix $\rho_X$  is defined as $F(\a)=\Tr\,\rho_X \a$.      
The function $F(\a)$ has three key properties, already introduced in section \ref{spins},  which characterize what is known as a ``state''
on an algebra:
\begin{itemize}\item{  it is linear in $\a$, namely $F(\lambda \a+\mu \b)=\lambda F(\a)+\mu F(\b)$, for $\a,\b\in\A_X$, $\lambda,\mu\in \C$,} \item{ it is normalized
to $F(1)=1$,} \item{  it is positive, in the sense that  $F(\a^\dagger \a)\geq 0$ for all $\a\in\A_X$.} \end{itemize} 
So a density matrix $\rho_X$ for the subsystem $X$ is a state on the algebra $\A_X$ consisting of all operators on the Hilbert space $\H_X$ of
subsystem $X$.

In quantum field theory, there is no way to associate a Hilbert space to a local region in spacetime.   The best one can do is to associate to an  open set $\W\subset M$  a corresponding algebra of operators $\A_\W$.
The algebras $\A_\W$ obey physically motivated axioms, such as a condition associated to causality: 
$\A_\W$ and $\A_{\W'}$ commute (in the $\Z_2$-graded sense if fermions are present) if the regions $\W$ and $\W'$ are spacelike separated.\footnote{See for example
section 5.3.1 of the article by Brunetti and Fredenhagen in \cite{BF} for a useful statement of axioms in the context of curved spacetime.  In
Minkowski space, a standard reference is \cite{HaagBook}.} 
We will discuss in more detail presently the construction of $\A_\W$.   For now, we just  note  a difference
between a local region in quantum field theory and 
 a subsystem in ordinary quantum mechanics:  the algebra $\A_\W$ is,
 in the von Neumann
algebra language,  of Type III, not Type I \cite{ArakiQFT,Fred,HaagBook,Yng,WittenNotes}.   This will be explained in section \ref{Thermo},
in the analogous setting of quantum statistical mechanics.    The Type III nature of the algebras is actually the reason that there is no way to associate a Hilbert space
to a local region.  
If the algebra $\A_\W$ were of Type I, it would have an irreducible representation in a Hilbert space, and this Hilbert space would be
naturally associated to the open set $\W$ in spacetime.   However, 
a Type III algebra does not have an irreducible representation in a Hilbert space. Accordingly in quantum field theory,
there is no natural way to associate a Hilbert space $\H_\W$ to an open set $\W$.
All that we can really associate to the region $\W$  is the algebra $\A_\W$, not a preferred Hilbert space that it acts on.

There is also no good notion of a density matrix for the region $\W$, since this notion is not applicable for a Type III algebra.
But the notion of a state of the algebra
 $\A_\W$ -- a linear function obeying conditions that were stated previously -- does make sense.  This is the appropriate analog
 in quantum field theory of the notion of the density matrix of a subsystem in ordinary quantum mechanics.

With this in mind, what should quantum dynamics mean for a local region $\U\subset M$?   We assume as described earlier that $\U$
is the domain of dependence of some set $V$, which has a slight thickening to an open set $\V\subset \U$.   If we could associate a Hilbert space
to quantum fields on  $V$ or in $\V$, we could describe quantum dynamics in terms of state vectors.  
We would say that a quantum state that defines initial conditions on $V$ or in $\V$ actually determines the probabilities for measurement outcomes in the larger region $\U$.
 Since instead we only have
algebras of operators associated to $\V$ and $\U$, we have to say something similar in terms of algebras.
The necessary statement is just that in a quantum field theory, in this situation $\A_\U=\A_\V$, meaning that operators in region
$\U$ are actually equivalent to operators in region $\V$ (although simple operators in region $\U$ might correspond to rather complicated
operators in the smaller region $\V$).   Since $\A_\U=\A_\V$, a state of $\A_\V$ is automatically a state of $\A_\U$: in other words,
the data needed to predict  measurements  outcomes in region $\V$ (to the extent that such outcomes are predictable in quantum
mechanics) also suffice to predict measurement outcomes in the larger region $\U$.   The statement that $\A_\U=\A_\V$, together
with general axioms about the local algebras, such as those described in \cite{BF,HaagBook}, is the content of quantum dynamics
for a local region.   

Now let us discuss in more detail the definition of the algebra $\A_\W$ associated to an open set $\W$.    We assume to begin with
that $\W$ is contained in a closed universe $M$, or at least in some spacetime (such as Minkowski space) to which the quantum field
theory of interest associates a  Hilbert space $\H$. 
A standard approach to defining $\A_\W$  is as follows.  Let $\O(x)$ be a local operator of the 
quantum field theory in question, and $f$ a smooth function with compact support in $\W$.    Then $K_f=\exp(\i\int_\W\d^Dx \sqrt g f(x) \O(x))$
is  a bounded operator on $\H$.   $\A_\W$ is then defined as the von Neumann algebra generated by such operators on $\H$.   

The purpose
of introducing the global Hilbert space $\H$ was to make sure that expressions such as $K_f$ can be interpreted as Hilbert space operators
so that it is possible to take weak limits and define a von Neumann algebra.   
The von Neumann algebra language is useful because it makes possible simple statements such as $\A_\U=\A_\V$ (which would not hold
if we do not complete the algebras by taking weak limits).   However, this language has the drawback of not directly incorporating the operator
product expansion (OPE), which is an important statement of the locality of quantum field theory.  One approach to incorporating the OPE is to enrich the von Neumann algebra language with a further axiom which would imply the existence of an OPE \cite{Bost}.   Another approach is to
state axioms for quantum field theory in curved spacetime that directly incorporate the OPE \cite{HW}.   In some approaches, the notion of a 
state has to be refined to incorporate the expected short distance singularities of the quantum field theory under study, but it is preferable if
(as in \cite{HW}) the expected short distance behavior can be built into the structure of the algebra.   The best treatment of such questions is not entirely clear in the author's opinion, and in this article we will not adopt any particular point of view.

Before discussing an open universe, we need a few more facts about local regions in a closed universe $M$.
Let us specialize to the case of an open set that is a local region $\U\subset M$, or a smaller open set $\V\subset\U$, as described earlier.
Thus $\U$ or $\V$ is globally hyperbolic, with a Cauchy hypersurface $V$ that is an open subset of a Cauchy hypersurface $S$ of $M$.
There are many ways to embed $\U$ or $\V$ in some other closed universe $M'$, such that $V$ is an open subset of a Cauchy hypersurface
$S'$ of $M'$.   For this, we simply modify $S$ outside of $V$, or $M$ outside of $\U$ or $\V$.     
However, $\A_\U$ and $\A_\V$ depend only on $\U$ and $\V$, and not on how $\U$ and $\V$ have been embedded in $M$.
This has been called ``the principle of locality'' \cite{Kay}.  For free field theory, the statement is a theorem.
In general, it is expected to hold because the algebras $\A_\U$ and $\A_\V$ are ultimately determined by operator product relations
in the regions $\U$ and $\V$, and these relations are entirely local in nature.   The principle of locality is analogous to what we learned about spin
systems in section \ref{spins}: there are many inequivalent representations of the algebra of an infinite collection of spins, but they are all equivalent
for any finite set of spins.

Given the principle of locality, we can  explain the meaning of quantum field theory in an open universe $M'$, now with a noncompact
initial value surface $S'$.   Let $V\subset S'$ be an open set that is small enough that it can be embedded in a closed
manifold $\t V$. Let $\U$ be the domain of dependence of $V$ in $M'$, and $\V\subset \U$ a small thickening of $V$.
Because $V$ can be embedded in the compact manifold $\t V$, $\U$ and $\V$ can be embedded in a closed universe $\t M$,
with $\t V$ as Cauchy hypersurface.  Then we can define algebras $\A_\U$ and $\A_\V$ associated to $\U$ and $\V$, and the
principle of locality says that these local algebras depend only on $\U$ and $\V$ and not on the embedding in $M'$.
The dynamical principle of a quantum field theory is now again $\A_\U=\A_\V$.   This, along with general axioms about the algebras
associated to open sets, is the content of quantum field theory in an open universe.

The principle of locality sheds light on what happens if we prefer to describe physics in an open universe using a Hilbert space.  The only problem with
doing so is that in an open universe there are many inequivalent choices of a Hilbert space
and generically there is no preferred way to choose one.   However, the principle of locality says that the  algebra $\A_\U$ of a local
region and the dynamical principle $\A_\U=\A_\V$ do not depend on the choice of a Hilbert space.      Because of the principle of locality, an observer interested in physics in a bounded portion of the universe may decide
that the choice of a global Hilbert space from among the myriad possibilities is irrelevant.   But one can pick whichever Hilbert space one wishes without
changing the predictions of a theory for local dynamics.

In an open universe that obeys special asymptotic conditions at spatial infinity, one can say more.    One particularly important case is an asymptotically
flat universe, that is a universe that is asymptotic at spatial infinity to Minkowski space. 
In the asymptotically flat case, we will assume that the theory of interest has a mass gap, for a reason explained in the next paragraph.
  Another important case is an asymptotically Anti de Sitter
(AAdS) universe, that is, a universe that is asymptotic at spatial infinity to Anti de Sitter space.  In the AAdS case, one assumes a boundary condition at
spatial infinity of the sort usually considered in the AdS/CFT correspondence.   An asymptotically flat or AAdS spacetime is time-independent
for modes near spatial infinity, so one has a natural separation
in positive and negative frequencies for modes near spatial infinity.   For any mode of sufficiently short wavelength, there is a natural separation
in positive and negative frequencies.  In short, for the modes that either have short wavelength or are located near spatial infinity -- which
means for all but finitely many modes -- there is a natural asymptotic separation in positive and negative frequencies.  Hence, an asymptotically flat or AAdS spacetime (with a mass gap in the asymptotically flat case) 
is similar to a closed universe:  there is a natural treatment asymptotically for all but finitely many
modes, and therefore we should expect that a natural Hilbert space will exist.   

Clearly, the argument as stated in the last paragraph is only heuristic, and I am not aware of precise theorems in the literature.   For a quantum field theory
with a mass gap $m$, the effects of spacetime curvature on the ground state are exponentially small in the limit that the radius of curvature is large compared to
$1/m$.  In such a case, in an asymptotically flat spacetime,
the separation in positive and negative frequencies becomes exponentially  precise  near spatial infinity, where the radius of curvature  diverges,
so there should be no difficulty.  In AAdS space with the usual sort of boundary conditions, there are no infrared issues and this case should also be safe.
The delicate case is a massless theory in an asymptotically flat spacetime.\footnote{The following was explained to me in this context by R. Wald.}   
For a free Maxwell field in Minkowski space, there is a natural quantization
that leads to a Hilbert space with a Poincar\'e invariant vacuum, but this quantization is not truly unique because of the possibility of ``soft hair'' \cite{Ashtekar};
after coupling to charged fields, usually assumed to be massive, the soft hair plays an important role in understanding infrared divergences, soft theorems,
and the memory effect \cite{Strominger}.   With massless charged fields, the non-uniqueness of quantization becomes truly essential; moreover
this has a very interesting analog for
General Relativity in an asymptotically flat spacetime \cite{SWa}.   One cannot expect quantum field theory in an asymptotically flat spacetime to be simpler
than quantum field theory in Minkowski space.   So in general, a straightforward Hilbert space  description of quantum field theory in asymptotically flat
spacetime is only available for massive theories or possibly for massless theories whose infrared behavior is such
that the $S$-matrix can be defined in a conventional Fock space.

\subsection{Non-Free Theories}\label{nonfree}

So far we have discussed  free theories.   We will add a few words on non-free theories.

One of the main criteria for successful quantization of a free theory is that correlation functions such as the two-point function
$\la\Psi|\phi(\vec x,t)\phi(\vec x',t')|\chi\ra$ should have standard short distance singularities.   A two-point function with  that property 
 is the essential input to perturbation theory.   So one would expect, modulo some well-understood issues of anomaly cancellation,\footnote{A theory which is well-defined in Minkowski space might have a gravitational anomaly which would cause perturbation theory in curved spacetime to fail.}
that weakly coupled theories make sense in perturbation theory when free theories do.    See \cite{Fre,Wal} for some rigorous results.

One would anticipate that the same is true nonperturbatively, for theories such as QCD.  If a theory exists perturbatively in curved spacetime, and nonperturbatively
in flat spacetime, one would expect that it works nonperturbatively in curved spacetime.   Unfortunately, not much is available in terms of rigorous
theorems, except for special models like two-dimensional conformal field theories.   That reflects the general mathematical difficulty of understanding
quantum field theory rigorously.  One would think that rigorous results for a superrenormalizable theory in curved spacetime might be relatively accessible,
but such results are not available.

\section{Quantum Statistical Mechanics and the Thermodynamic Limit}\label{Thermo}

\subsection{The Thermofield Double}\label{td}

In this section, we will consider a simpler problem that is actually somewhat analogous to quantum field theory in an open universe.
Instead of quantum field theory in curved spacetime, we consider quantum statistical mechanics at positive temperature $T=1/\beta$.   In finite
volume, there is no problem.   We just consider the density matrix $\rho=\frac{1}{Z}e^{-\beta H}$ acting on the Hilbert space
$\H$ of the system, where $Z=\Tr\,e^{-\beta H}$ is the partition function.   Importantly, for any conventional statistical
system (possibly a relativistic quantum field
theory, possibly a nonrelativistic system of some kind), the partition function converges in  finite volume.  Thermodynamic functions and 
thermal correlation functions can all be computed in finite volume, and they have a thermodynamic or large volume limit.   

What happens if one wants to describe the thermodynamic limit of a system directly in terms of operators acting on
a Hilbert space appropriate for an infinite volume system?    Here we run into a difficulty: there is no  separable Hilbert space
that contains all the typical states of a system at $T\not=0$, since fluctuations can occur throughout the whole infinite space.  
  For example, for a spin system on an infinite lattice, since each spin has a nonzero probability to fluctuate,
it takes uncountably many states to describe all the typical states at $T\not=0$. (We discuss some examples based on lattice spin systems
in section \ref{lss}.)    In a continuum field theory, there is actually no reasonable
definition of a nonseparable Hilbert space that contains all the typical thermal excitations in infinite volume.

There is a simple device that enables one to describe the thermodynamic limit of any quantum system  in a separable Hilbert space \cite{HHW}.   This is the
``thermofield double.''   Consider an ordinary quantum system $\T$ (possibly a relativistic field theory, possibly a lattice theory or something else) 
with Hilbert space $\H$  and Hamiltonian $H$.
In finite volume, the thermofield double is defined by introducing, roughly speaking,  two copies of the original
system,  
 which we will call the ``right'' and ``left'' copies, with Hilbert spaces $\H_r$ and $\H_\ell$, and Hamiltonians $H_r$ and $H_\ell$.
We think of $\H_r$ as the Hilbert space of the system $\T$ and $\H_\ell$ as an auxiliary second copy.
Actually, rather than being a second copy of $\H_r$, it is more canonical to define $\H_\ell$ to be the complex conjugate Hilbert space of $\H_r$.   That amounts
physically to saying that the left system is a time-reversal conjugate of the right system (rather than being an identical second copy).  The fact
that $\H_\ell$ is the complex conjugate of $\H_r$ ensures that the combined Hilbert space  $\H_\TFD=\H_\ell\otimes \H_r$ of the two systems can  be viewed canonically
as the space of linear operators acting on $\H_r$ (or $\H_\ell$).  In expositions of the thermofield double,  time-reversal symmetry is often assumed,
in which case the left and right systems are equivalent.

In finite volume, the thermofield double $\Psi_\TFD$ is simply a vector in $\H_\ell\otimes \H_r$, defined as follows.  Suppose that the Hamiltonian
acting on the original system has eigenstates $\psi_i$ with energies $E_i$.  Then in finite volume, $\Psi_\TFD$ is defined as
\be\label{tomigo} \Psi_\TFD =\frac{1}{\sqrt Z} \sum_i \exp(-\beta E_i/2) |\bar i\ra_\ell\otimes |i\ra_r,\ee
Here $|i\ra_r$ is the $i^{th}$ energy level of the right system, with energy $E_i$; $|\bar i\ra_\ell$ is its time-reversal conjugate in the left system; and 
 $Z$ is the partition function  of the right or left system at temperature $1/\beta$.
 The point of this formula is that the density matrix of the right or left system is the usual thermal density matrix of that system:
\be\label{rumo}\rho_r=\Tr_{\H_\ell}|\Psi_\TFD\ra\la\Psi_\TFD| =\frac{1}{Z} \sum_i e^{-\beta E_i}|i\ra_r\la i|_r=\frac{1}{Z}e^{-\beta H_r}.\ee
So the thermofield double is a ``purification'' of a thermal density matrix.  

An equivalent statement is the following.  If we view $\H_\ell\otimes \H_r$ as the space of linear operators acting on $\H_r$, then $\Psi_\TFD=\rho_r^{1/2}$.  
To exploit this fact, let  $\A_{r,V}$ and $\A_{\ell,V}$  be the algebras of all operators acting on $\H_r$ or $\H_\ell$, respectively, in finite volume $V$. Viewing a state
vector $\Psi\in \H_\ell\otimes \H_r$ as a matrix acting on $\H_r$,  an element $\a_r\in\A_{r,V}$ acts on $\Psi$ by $\Psi\to \a_r\Psi$, and an element $\a_\ell\in\A_{\ell,V}$ 
acts on $\Psi$ by $\Psi\to\Psi\a_\ell^\tr$ (where $\a_\ell^\tr$ is the transpose of $\a_\ell$).     If $\Psi$, viewed as a matrix acting on $\H_r$, is invertible,
then every matrix acting on $\H_r$ is of the 
form   $\a_r\Psi$ and of the form  $\Psi\a_\ell^\tr$ 
for some unique
$\a_r\in \A_{r,V}$ and for some unique $\a_\ell\in\A_{\ell,V}$.   In particular, $\Psi_\TFD=\rho_r^{1/2}$ is
invertible, and therefore every vector in $\H_\TFD$ can be obtained from $\Psi_\TFD$ by acting with $\A_{r,V}$, or by acting with $\A_{\ell,V}$.  
The fancy way to say this is that if $\Psi$ is invertible, then it corresponds to a cyclic separating vector\footnote{ \label{cc}  In general, if an algebra $\A$ acts on a Hilbert space $\H$, then a vector $\Psi\in\H$ is said to be separating for $\A$ if
the condition $\a\Psi=0$, $\a\in\A$ implies that $\a=0$, and it is said to be cyclic for $\A$ if vectors $\a\Psi$, $\a\in \A$ are dense in $\H$.} for the
algebras $\A_{r,V}$ and $\A_{\ell,V}$.

 \begin{figure}
 \begin{center}
   \includegraphics[width=3.5in]{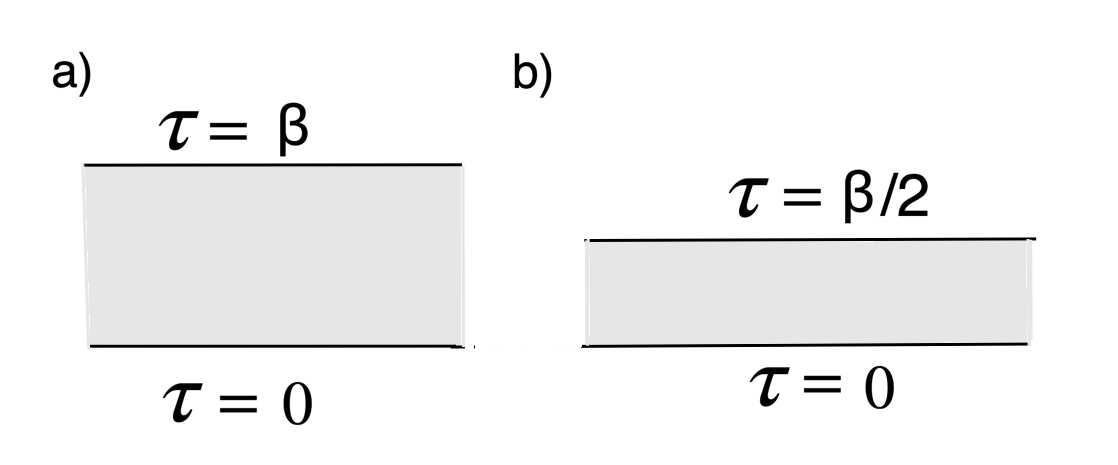}
 \end{center}
\caption{\small  (a) A thermal density matrix $\rho=\frac{1}{Z}e^{-\beta H}$ can be computed by a path integral on the strip
$0\leq \tau\leq \beta$.  The horizontal direction in the picture represents the ``spatial''  manifold (or lattice) $W$.  (b) The thermofield double state $\Psi_\TFD=\rho^{1/2}$ can be computed by a path integral on a strip
of half the width, $0\leq \tau\leq \beta/2$.   Operators of the ``right'' or ``left'' copy are inserted, respectively, on the boundaries at $\tau=0$ and $\tau=\beta/2$. .  \label{TFDpicture}}
\end{figure} 

We want to take the infinite volume limit of this construction.  As already noted, the left and right Hilbert spaces $\H_r$ and $\H_\ell$
do not have large volume limits, and the formula (\ref{tomigo}) is not meaningful in the thermodynamic limit, because $Z=\infty$ in this limit and
the sum would contain uncountably many terms.
 But it is possible to define a Hilbert space $\H_\TFD$ that contains $\Psi_\TFD$ and
does have a large volume limit.   To explain this, we first review
a few standard facts about finite volume.  We describe these facts in the familiar language of path integrals, though this is not strictly necessary.
 Let $W$ be the spatial manifold (or lattice) on which we want
to formulate theory $\T$. In the thermodynamic limit, $W$ will be $\R^{D-1}$ for a continuum field theory or an infinite lattice   $\Z^{D-1}$ for
a lattice system.    The thermal density matrix $\rho_r=\frac{1}{Z}e^{-\beta H}$ can be computed in finite volume by a path integral on $W\times I$,
where $I$ is the interval $0\leq \tau\leq \beta$ (fig. \ref{TFDpicture})
 The factor of $1/Z$ means that from the energy density $T_{00}$, whose integral is the Hamiltonian $H_r=\int\d^{D-1}x \sqrt g T_{00}$, we should
 subtract a constant, the free energy density at temperature $1/\beta$.     With this subtraction, the path integral, in finite volume, computes a normalized density matrix
 $\rho_r$, satisfying
 $\Tr\,\rho_r=1$.   To construct the thermofield double state $\Psi_\TFD=\rho_r^{1/2}$, we simply do the same path integral (with the same subtraction)
 on a strip of half the width, $0\leq \tau\leq \beta/2$.   Given an operator $\a_r\in \A_{r,V}$, to construct a state obtained
 by acting with $\a_r$ on $\Psi_\TFD$, we just insert $\a_r$ in the path integral at $\tau=0$; similarly, to act with $\a_\ell\in \A_{\ell,V}$ on $\Psi_\TFD$,
 we insert $\a_\ell^\tr$ at $\tau=\beta/2$.   Inner products between these states are computed by gluing together two path integrals on the strip of
 width $\beta/2 $ to make a cylinder of circumference $\beta$.    For example,
 an inner product $\la\b_r\Psi_\TFD|\a_r\Psi_\TFD\ra=\la\Psi_\TFD|\b_r^\dagger\a_r \Psi_\TFD\ra$ is computed by a path integral on the cylinder with insertion
 of $\b_r^\dagger \a_r$ at $\tau=0$.   This quantity is just the expectation value of $\b_r^\dagger \a_r$ in the thermal ensemble:
 \be\label{waggo}\la\b_r\Psi_\TFD|\a_r\Psi_\TFD\ra=\la\Psi_\TFD|\b_r^\dagger\a_r     |   \Psi_\TFD\ra=\Tr\,\rho_r \b_r^\dagger\a_r
 =\frac{1}{Z}\Tr_{\H_r }\exp(-\beta H_r) \b_r^\dagger \a_r.\ee Such  thermal expectation values have a thermodynamic
 limit.\footnote{In taking the thermodynamic limit, we keep the operators fixed, and hence acting on degrees of freedom in a bounded region, while
 the volume goes to infinity.}   Since the density operator $\rho_r$ is strictly positive, one has
 \be\label{toldo}\la\a_r\Psi_\TFD|\a_r\Psi_\TFD\ra= \Tr_{\H_r}\rho_r \a_r^\dagger \a_r>0\ee
 for all $\a_r\not=0$.
  Likewise, for $\a_\ell\not=0$, we can construct states $\a_\ell\Psi_\TFD$ by acting with $\a_\ell^\tr$ at $\tau=\beta/2$.  
  Inner products between these states are obviously given by a formula like eqn. (\ref{waggo}) which again has a thermodynamic limit:
  \be\label{napgo} \la\b_\ell\Psi_\TFD|\a_\ell\Psi_\TFD\ra=\frac{1}{Z}\Tr_{\H_\ell }\exp(-\beta H_\ell) \b_\ell^\dagger \a_\ell.\ee
   Inner products
  $\la\a_\ell\Psi_\TFD|\a_r\Psi_\TFD\ra$ are defined by inserting the operators $\a_r$ and $(\a_\ell^\tr)^\dagger=\a_\ell^*$ (here $\a_\ell^*$ is the complex conjugate of $\a_\ell$; note that because of the complex conjugation, it is naturally an operator on $\H_r$)
  at $\beta=0$ and $\beta=\tau/2$, respectively.  In operator terms, such an inner product is 
  \be\label{moldo} \la\a_\ell\Psi_\TFD|\a_r\Psi_\TFD\ra=\frac{1}{Z}\Tr_{\H_r}\, \a_r e^{-\beta H_r/2} \a_\ell^* e^{-\beta H_r/2}. \ee
Such an inner product also has a thermodynamic limit.   Reflection positivity says that for any $\a_\ell$, there is an $\a_r$ that makes this inner product
nonzero.  We simply choose $\a_r$ to be obtained from $\a_\ell$ by a reflection of the thermal circle.  

In finite volume, the states that appeared in this construction are simply states in $\H_\ell\otimes \H_r$.   In the infinite volume limit,
this is no longer true.  But now we can describe a Hilbert space $\H_\TFD$ that does have a thermodynamic limit.
First we need to pick an algebra of operators that is supposed to act on $\H_\TFD$.   As in section \ref{spins}, we at least want to include
all operators that act on a bounded region of space.  Ultimately we can take limits and include operators whose action is not restricted to a bounded region, but
to know what limits to allow, we first have to know what Hilbert space the operators are supposed to act on.   So we define algebras
$\A_{r,0}$ and $\A_{\ell,0}$ consisting of bounded operators acting on a bounded region in the right system or the left system, respectively.  Now at a minimum we want
$\H_\TFD$ to contain a vector that we call $\Psi_\TFD$ which will be an infinite volume limit of the finite volume thermofield double state.   In addition, we want the algebras $\A_{r,0}$ and $\A_{\ell,0}$ to act on
$\H_\TFD$.   So we say that  for every $\a_r\in\A_{r,0}$, $\H_\TFD$ contains a vector that we call $\a_r\Psi_\TFD$.  $\A_{r,0}$ acts on these
states in an obvious way: $\a_r(\b_r\Psi_\TFD)=(\a_r\b_r)\Psi_\TFD$. We postulate that the inner products
between these states are given by the thermodynamic limit of the thermal correlation functions in eqn. (\ref{waggo}).     Because of eqn. (\ref{toldo}), this
set of states makes up a pre-Hilbert space $\H_{\TFD,0}$, satisfying all conditions of a Hilbert space except completeness.   We take the Hilbert space
completion and get the thermofield double Hilbert space $\H_\TFD$.   We can now also complete $\A_{r,0}$ to a von Neumann algebra $\A_r$ 
that acts on $\H_\TFD$ by including
weak limits, as in section \ref{spins}.

Obviously, we could have carried out this process starting with $\A_{\ell,0}$ instead of $\A_{r,0}$.    One might worry that this would lead to a different
thermofield double Hilbert space $\widetilde\H_\TFD$, with an action of a left algebra $\A_\ell$ instead of $\A_r$.   The reason that this does not happen
is that the formula eqn. (\ref{moldo}) for inner products between states $\a_\ell\Psi_\TFD$ and states $\a_r\Psi_\TFD$ also has a thermodynamic
limit and  implies that $\a_\ell\Psi_\TFD$ can be viewed as a vector in the same Hilbert space $\H_\TFD$ that we defined starting with states $\a_r\Psi_\TFD$.

The upshot is to define a Hilbert space $\H_\TFD$ with an action of completions $\A_\ell$ and $\A_r$ of $\A_{\ell,0}$ and $\A_{r,0}$.   
Inner products between vectors of the form $\a_\ell\Psi_\TFD$ and/or $\a_r\Psi_\TFD$ are defined via the thermodynamic limits of eqns.
(\ref{waggo}), (\ref{napgo}), and (\ref{moldo}).  All expected relations among inner products between Hilbert space vectors are satisfied, because
we are simply taking the thermodynamic limits of formulas that in finite volume do represent inner products between Hilbert space vectors.  
Moreover, in terminology that was defined in footnote \ref{cc}, $\Psi_\TFD\in\H_{\TFD}$ is always a cyclic separating vector for the algebra $\A_r$ (or $\A_\ell$).
     
     The procedure by which we constructed the thermodynamic limit of $\H_\TFD$ is a special case of the Gelfand-Naimark-Segal (GNS) construction
     in which the input is an algebra $\A$ and a state on the algebra, that is a complex-valued linear function $\a\to F(\a)$ such that  $F(1)=1$ and $F(\a^\dagger\a)>0$ for all $\a\not=0$,
     and the output is a Hilbert space $\H$ with an action of $\A$ and a cyclic separating vector $\Psi$.   (Closely related  is the Wightman
     Reconstruction Theorem of axiomatic field theory, in which a Hilbert space is reconstructed from the correlation functions of a local field.)
     The GNS construction proceeds as follows.  One formally associates a vector $\Psi$ to the identity element $1\in \A$, and to every $\a\in \A$
     one associates a vector $\a\Psi$.  
     The action of $\A$ on this set of states
     is defined in the obvious way by $\a(\b\Psi)=(\a\b)\Psi$, and inner products are defined by $\la\a\Psi|\b\Psi\ra=F(\a^\dagger \b)$.  This gives a pre-Hilbert space $\H_0$ with an action of $\A$, and one completes it to a Hilbert space
     $\H$.  From the definitions, it is immediate that $\A$ acts on $\H$ with $\Psi$ as a cyclic separating vector.  Clearly the definition of $\H_\TFD$ is
     the GNS construction applied to the case that $F(\a)$ is the expectation value of $\a$ in the thermodynamic limit.     

\subsection{Surprises In The Thermodynamic Limit}\label{surprises}

So by going to the thermofield double, we can describe the infinite volume limit of quantum statistical mechanics in a separable Hilbert space $\H_\TFD$.
But there are a few surprises.

First of all, evidently $\A_r$ does not act irreducibly on $\H_\TFD$, since it commutes with $\A_\ell$, and
vice-versa. In fact  $\A_\ell$ and $\A_r$ are algebras of a possibly unfamiliar kind: they are generically von Neumann algebras of Type III.
What characterizes algebras of Type II or Type III, as opposed to the more familiar algebras of Type I, is roughly that entanglement is not just a property of the states
that the algebra acts on but is built into the structure of the algebra.
In finite volume $V$, the entanglement entropy of the state $\Psi_\TFD\in\H_\ell\otimes \H_r$ is the same as the thermodynamic entropy of the thermal
density matrix $\rho_r$ or $\rho_\ell$ on one factor.  In particular, it is proportional to $V$ and diverges  in the thermodynamic limit.   Thus in effect $\Psi_{\TFD}$ in the thermodynamic
limit describes a state with an infinite entanglement between the two systems.     But every state in $\H_\TFD$ looks at spatial infinity like $\Psi_\TFD$ and so every state
has the same divergent entanglement entropy.   This is different from the usual situation for a bipartite quantum system $XY$ with  Hilbert space $\H_{XY}=\H_X
\otimes \H_Y$  where $\H_X$
and $\H_Y$ are of infinite dimension.   In such a case, a vector $\Psi_{XY}\in \H_{XY}$ {\it could} have infinite entanglement entropy,
but there are also vectors in $\H_{XY}$ with finite or zero entanglement entropy.  By contrast, in the infinite volume limit, $\H_\TFD$ is not a tensor product  
and no vector in it can be interpreted to have finite entanglement entropy; rather,
the infinite entanglement
entropy between the left and right systems is encoded in the structure of the algebras $\A_\ell$ and $\A_r$.
We will discuss this more thoroughly in section \ref{lss} with simple examples.

A second unusual property of the infinite volume thermofield double comes to light if we think about real time thermal physics.
In finite volume, to study real time thermal physics, we define  the time-dependence of operators in a familiar way: 
\begin{align}\label{standef}  \a_r(t)&=\exp(\i H_r t)\a_r \exp(-\i H_r t)\cr
\a_\ell(t)&=\exp(\i H_\ell t)\a_\ell \exp(-\i H_\ell t).\end{align}
 Then we define real time correlation functions such as
\be\label{toffo} \la \a_r(t_1)\b_r(t_2)\ra =\Tr\,\rho_r  \a_r(t_1)\b_r(t_2). \ee
These functions have a thermodynamic limit.    How do we describe them using the thermofield double?   A subtlety arises here: the Hamiltonians $H_\ell$
and $H_r$ of the left and right system do not make sense as operators on the infinite volume $\H_\TFD$.   This is so because of thermal fluctuations.
In volume $V$, the typical energy in the thermal ensemble is of course proportional to $V$, so the Hamiltonian diverges for $V\to \infty$.   
That in itself is not a problem, because we can ``renormalize'' $H_\ell$ and $H_r$ by subtracting a multiple of $V$ to get an operator whose expectation value
vanishes for $V\to\infty$.   The problem comes from the thermal fluctuations.   
    The typical fluctuations in $H_\ell$ and $H_r$ in volume $V$ are of order $V^{1/2}$, which also diverges for $V\to\infty$.  Even after subtracting constants
    from $H_\ell$ and $H_r$ to set their expectations values in the thermal ensemble to zero, the divergent fluctuations remain, and imply that $H_\ell$ and $H_r$
    do not have limits as operators on $\H_\TFD$.
    
    So how are we going to define real time correlation functions?   The answer is that the operator $\h H=H_r-H_\ell$ does have a thermodynamic limit as an
    operator on $\H_\TFD$.    This results from the fact that in finite volume $\h H$ annihilates $\Psi_\TFD$, as is evident from the definition (\ref{tomigo}).
    With the aid of $\h H$, we can describe real time thermal physics.   We define
    \begin{align}\a_r(t)&=\exp(\i \h H t)\a_r \exp(-\i \h H t) \cr
                        \a_\ell(t)&=\exp(-\i \h H t)\a_\ell \exp(\i\h H t). \end{align}
     The motivation for these definitions is that in finite volume they agree with the standard definitions  (\ref{standef}), since $[H_\ell,\a_r]=[H_r,\a_\ell]=0$.
 In terms of these definitions, the thermodynamic limit of real time correlation functions is given by formulas such as
 \be\label{omgon}\la\a_{\ell,r}(t_1)\b_{\ell,r}(t_2)\ra=\la\Psi_\TFD|a_{\ell,r}(t_1)b_{\ell,r}(t_2)|\Psi_\TFD\ra,\ee where the subscripts $\ell,r$ mean that the operators can
belong to  either $\A_\ell$ or $\A_r$.
     
     Since the operator $\exp(\i \h H t)$ is not   contained in either $\A_\ell$ or $\A_r$, it follows that time translations are a group of outer automorphisms of
     $\A_\ell$ and of $\A_r$.   Now consider the following question.   Suppose that one has access to the Hilbert space $\H_\TFD$, the state $\Psi_\TFD$,
     and the algebra $\A_r$, but one does not know how this data arose by taking a thermodynamic limit.
     Just from $\H_\TFD$, $\Psi_\TFD$, and $\A_r$, 
       how would one identify which outer automorphism group of $\A_r$ corresponds to time translations?   The answer to this
     question is given by Tomita-Takesaki theory.   Here we will just state a few facts, leaving the reader to look elsewhere (for example, sections 3 and 4 of
     \cite{WittenNotes}) for
     a detailed explanation.  
      In Tomita-Takesaki theory, to a Hilbert space $\H$, an algebra $\A$ of operators on $\H$ and a state
     $\Psi\in\H$ that is cyclic and separating\footnote{See footnote \ref{cc} for the meaning of ``cyclic and separating.''}  for $\A$, one associates 
  a modular operator $\Delta_\Psi$.      For example, if $\H=\H_{XY}=\H_X\otimes\H_Y$ 
  is the Hilbert space
  of a bipartite system $XY$, $\A=\A_X$ is the algebra of operators on $\H_X$, and $\Psi=\Psi_{XY}$ is a vector in $\H_{XY}$,  then the cyclic separating
condition is that the reduced density matrices $\rho_X$ and $\rho_Y$ of $\Psi_{XY}$ are invertible, and in this case the modular operator is $\Delta_\Psi=\rho_X\otimes \rho_Y^{-1}$.
(See eqn. (4.26) in \cite{WittenNotes}, for example.)   In finite volume,  these conditions are satisfied by
$\H=\H_\TFD=\H_\ell\otimes \H_r$, $\Psi=\Psi_\TFD$, $\A=\A_r$, $\rho_r=\frac{1}{Z}e^{-\beta H_r}$, $\rho_\ell=\frac{1}{Z}e^{-\beta H_\ell}$.    So
$\Delta_{\Psi_\TFD}=\rho_r\otimes \rho_\ell^{-1}=\exp(-\beta \h H)$.   Though $\rho_\ell$ and $\rho_r$ do not have a thermodynamic
limit, $\Delta_{\Psi_\TFD}$ and $\h H$ do have such a limit and 
the formula $\Delta_{\Psi_\TFD} =\exp(-\beta \h H)$ is valid in the thermodynamic limit.
 So we can express $\h H$ in terms of the modular operator: $\h H=-\frac{1}{\beta }\log \Delta_{\Psi_\TFD}$.

\subsection{Examples From  Spin Systems}\label{lss}

Here we will consider some concrete examples of the thermofield double construction  for a system with an infinite
number of degrees of freedom.\footnote{Some of the following was described from a different point of view in section 6 of \cite{WittenNotes}.}   
As in section \ref{spins}, we consider a countably infinite set of qubits, which we label by a positive integer $n$, making the ``right'' system.   
To study this system via the thermofield
double, we introduce another infinite collection of qubits, the ``left'' system.  We write $\H_{r,N}$ for the Hilbert space of the first $N$ qubits
of the right system, and $\H_{\ell,N}$ for the analog for the left system.

We will consider extremely simple Hamiltonians of the form $H=\sum_{n=1}^\infty  H_n$, where $H_n$ is a single-qubit Hamiltonian.    One may be surprised
that this case is rich enough to be interesting.   In fact the various von Neumann algebras\footnote{To be more precise, the  ``hyperfinite'' von Neuman
algebras all arise this way.   A hyperfinite algebra is one that can be approximated by a finite-dimensional matrix algebra.  Hyperfinite algebras are usually the
ones that arise in physics, because infinite constructions in physics are usually limits of finite constructions.   The classification of von Neumann algebras is much more
complicated in the non-hyperfinite case.}
can all arise from such a construction and some
of them were first discovered in this way \cite{VN,Powers,ArakiWoods}.

The simplest Hamiltonian of all is $H=0$, so we will start with that case.   Let us write $|i\ra_{n,r}$, $i=1,2$ 
for the two states of the $n^{th}$ qubit of the  ``right'' system, 
and $|\bar i\ra_{n,\ell}$, $i=1,2$ for the time-reversed states of the $n^{th}$ qubit of the ``left'' system.   The analog of defining the thermofield double
state first in finite volume, as in eqn. (\ref{tomigo}), is to define the thermofield double state first for the first $N$ qubits of the right system, together with their
partners in the left system.   With $H=0$, the thermofield double state $\Psi_\TFD\in \H_{r,N}\otimes \H_{\ell,N}$  is just a completely entangled state of $N$ pairs of qubits:
\be\label{wacko}\Psi_{\TFD}=\frac{1}{\sqrt{Z_N}}\bigotimes_{n=1}^N \sum_{i=1,2}|i\ra_{n,r}\otimes |\bar i\ra_{n,\ell},~~~Z_N=2^{N}. \ee
Arranging the four states $|i\ra_{n,r}\otimes |j,\ra_{n,\ell}$, $i,j=1,2$  in a $2\times 2$ matrix in a fairly obvious way, we can write
\be\label{zacko} \Psi_\TFD=\frac{1}{\sqrt {Z_N}}\bigotimes_{n=1}^N\begin{pmatrix}1& 0\cr 0& 1\end{pmatrix}.\ee

Now let $\A_{r,0}$ be the algebra of all operators on the right system that act nontrivially on only finitely many qubits.     If $\a_r\in \A_{r,0}$, then for sufficiently large $N$,
 we can view $\a_r$ as an operator on $\H_{r,N}\otimes \H_{\ell,N}$ and define
$F(\a_r)=\la\Psi_\TFD|\a_r|\Psi_\TFD\ra$.   For given $\a_r$, this definition does not depend on $N$, once $N$ is large enough 
so that $F(\a_r)$ is defined.   So  $F(\a_r)$ is well-defined for all $\a_r\in \A_{r,0}$.
It clearly  is linear in $\a_r$ and satisfies $F(\a_r^\dagger\a_r)>0$ for $\a_r\not=0$, so it defines a state on the algebra $\A_{r,0}$.  
So we can invoke the GNS construction, as described in section \ref{td}, and construct the thermofield double Hilbert space $\H_\TFD$, with an action of
$\A_{r,0}$,  and  containing the distinguished  thermofield double state $\Psi_\TFD\in\H_\TFD$. 

By including weak limits, we can take the closure of $\A_{r,0}$ to get an algebrs $\A_r$ of bounded operators that acts on $\H_\TFD$.   This is an algebra of
an unfamiliar kind -- a Type II von Neumann algebra.   In fact, it is a Murray-von Neumann factor\footnote{A factor is a von Neumann
algebra whose center consists only of the complex scalars. }  of Type II$_1$ \cite{MvN}. 
To see that the algebra $\A_r$ is of an unfamiliar type, note that the function $F(\a_r)$ satisfies
\be\label{zombo} F(\a_r \b_r)=F(\b_r\a_r).  \ee
Thus, this function has the algebraic properties of a trace, so we will denote it as one: $F(\a_r)=\Tr\,\a_r$.   Moreover, this trace is defined (and  finite) for all elements of the algebra;
in particular, $\Tr\,1=1$.    By contrast, the obvious example of an infinite dimensional von Neumann algebra is the algebra $\A$ of all bounded operators on a
separable Hilbert space $\H$.   This algebra, which is said to be of Type I$_\infty$,
 does have a trace $\Tr$, but the trace is not defined for all elements of $\A$ (only for those that are ``trace class'') and
in particular  in a Type I$_\infty$ factor, $\Tr\,1=+\infty$.   So the algebra $\A_r$ is something new.

Intuitively, since the Type II$_1$ factor $\A_r$ was constructed by considering an infinite collection of maximally entangled qubit pairs,
an infinite amount of entanglement is built into the structure of this algebra.
A  Type II$_1$ algebra  does not have an irreducible representation in a Hilbert space.    
We constructed a  representation of the Type II$_1$ algebra
 $\A_r$  on the Hilbert space $\H_\TFD$, but this representation is far from irreducible, since the action of $\A_r$ on $\H_\TFD$ commutes with the action of another Type II$_1$ algebra $\A_\ell$
that is  defined similarly, starting with operators that act only on finitely many qubits of the left system and taking limits.    The algebras $\A_r$ and $\A_\ell$
commute and in fact they are each other's commutants.\footnote{The commutant of a von Neumann algebra $\A$ acting on a Hilbert space $\H$ is the algebra $\A'$
that consists of all bounded operators on $\H$ that commute with $\A$.}   As this example illustrates, when  a Type II$_1$ algebra acts on a Hilbert space, its commutant is   another
Type II algebra.

Murray and von Neumann classified the Hilbert space representations of a hyperfinite Type II$_1$ algebra.  They are labeled up to isomorphism by a regularized ``dimension'' $x$, which is a non-negative real number or $\infty$.   $\H_\TFD$ corresponds to the case $x=1$.   To construct representations with $x<1$, we use projection operators in $\A_\ell$.    For example, for $1\leq k\leq 2^N$, 
pick a $k$-dimensional subspace $W\subset \H_{\ell,N}$, and let $\Pi:\H_{\ell,N}\to W$ be the orthogonal projection operator of $\H_{\ell,N}$ onto $W$ (tensored with the identity operator on all other qubits).   The image of $\Pi$ in
$\H_\TFD$ is a representation of $\A_r$ with  $x=k/2^N$.  All values $x<1$ are limits of this with $k,N\to\infty$.   Since $x$ is additive when one takes the
direct sum of two representations, any representation is the direct sum of a representation with $x<1$ and a number of copies of $\H_\TFD$.   
Roughly, the value of $x$ for a Hilbert space representation $\RR$ of $\A_r$ is defined as the trace of the identity element of $\A_r$ in the representation $\RR$.
Some more detail is needed to make this precise, but anyway the existence of inequivalent representations of $\A_r$ is related to the fact that $\A_r$ has a trace.

   The other hyperfinite von Neumann algebra of Type II is a Type II$_\infty$ algebra.   It can
be obtained as the tensor product of a Type II$_1$ algebra with a Type I$_\infty$ algebra of all bounded operators on a separable Hilbert space.
It has a trace, which, because of the Type I$_\infty$ factor, is not defined for all elements of the algebra.   

Repeating this construction with a non-zero Hamiltonian, we can get the other hyperfinite von Neumann algebras.   Again 
with $H=\sum_{n=1}^\infty H_n$, the next
simplest case is to take 
\be\label{nextcase}H_n=\begin{pmatrix} 0&0\cr 0&E \end{pmatrix},\ee all with the same constant $E$.   The thermofield double state, in the notation of (\ref{zacko}),
is
\be\label{ozacko} \Psi_\TFD=\frac{1}{\sqrt {Z_N}}\bigotimes_{n=1}^N  \begin{pmatrix}1& 0\cr 0& e^{-\beta E/2}\end{pmatrix}, ~~~~Z_N=(1+e^{-\beta E})^N.\ee
In the large $N$ limit, we can again define a state on the algebra $\A_{r,0}$ by $F(\a_r)=\la\Psi_\TFD|\a_r|\Psi_\TFD\ra$.   The GNS construction gives
the thermofield double Hilbert space $\H_\TFD$, with an action on $\H_\TFD$ of an algebra $\A_r$ that is a closure of $\A_{r,0}$, and with a cyclic separating state $\Psi_{\TFD}$.

The state $\Psi_{\TFD}$ in this example describes  an infinite collection of qubit pairs all with the same non-maximal entanglement.   So 
$\A_r$ is again an algebra that encodes an
infinite amount of entanglement.  What is different for $E\not=0$ is that generically $F(\a_r b_r)\not= F(\b_r\a_r)$.   So the function $F$ is no longer
a trace, and indeed for $E\not=0$, $\A_r$  does not have a trace.   The algebra $\A_r$ is called an algebra of Type III$_\lambda$ with $\lambda=e^{-\beta E/2}$.
An algebra of this type was originally constructed in precisely the way that we have described \cite{Powers}.

In almost any real problem in quantum statistical mechanics, the algebra $\A_r$ that acts on the thermofield double is instead a von Neumann algebra
of Type III$_1$.   The original construction of such an algebra \cite{ArakiWoods} was based on again taking $H=\sum_{n=1}^\infty H_n$ to be a sum of
single-qubit Hamiltonians, but with different $H_n$:
\be\label{nextcasse}H_n=\begin{pmatrix} 0&0\cr 0&E_n \end{pmatrix}.\ee   In this general form, we will call the model the Araki-Woods model.
As long as $\lim_{n\to\infty } E_n\not=\infty$, there is an infinite amount of entanglement between the left and right systems, and the algebra
$\A_r$ that acts on $\H_\TFD$ will not be of Type I.  (The case $\lim_{n\to\infty} E_n=\infty$ is discussed in section \ref{ht}.)
 Except in rather special cases, $\A_r$ is a new type of  algebra that is said to be of Type III$_1$.   For example, if the $E_n$ take
two values $E$ and $E'$, each with infinite multiplicity, then, for generic $E$ and $E'$, the algebra $\A_r$  is of Type III$_1$.
This is again an algebra without a trace.

Unlike a Type II$_1$ algebra, a Type III algebra only has one  representation in a Hilbert space, up to 
isomorphism.    In the construction that we have just
described, the representation $\H_\TFD$ of
the Type III$_1$ algebra $\A_r$  is far from irreducible, since  the commutant of $\A_r$ is another Type 
III$_1$ algebra $\A_\ell$ that can be defined in the same way.
We can again choose a projection operator $\Pi\in\A_\ell$, and the image of $\Pi$ is a subspace  
$\Pi\H_\TFD\subset \H_\TFD$ that provides a representation of $\A_r$.
But as representations of $\A_r$, $\H_\TFD$ and $\Pi\H_\TFD$ are equivalent, even though one might naively 
think that the second one is smaller.   For Type II$_1$, the existence of the invariant $x$ that distinguishes
representations of different ``size'' is tied to the existence of a trace; for Type III, there is no trace 
and any two nontrivial representations are isomorphic.

Following is a summary of some facts about (hyperfinite) von Neumann algebras.   In all of these statements, 
we assume that the algebra considered is a ``factor''
(its center consists only of complex scalars).   Algebras that are not factors can be constructed in a relatively simple way from factors.  

(1) A Type I algebra $\A$ has an irreducible representation in a Hilbert space $\RR$. Its commutant in this representation consists only of its center, the
complex scalars.  This representation is essentially unique in the sense that any representation of $\A$ is a direct sum of copies of $\RR$;
differently put, any representation is of the form $\RR\otimes \Q$, where $\Q$ is some
other Hilbert space on which $\A$ acts trivially.  
  $\A$ is said to be of Type I$_n$ if $\RR$ is $n$-dimensional,
and of Type I$_\infty$ if $\RR$ has countably infinite dimension.

(2) A Type II factor $\A$ has no irreducible representation in a Hilbert space; its commutant 
in any representation is another Type II factor.   There are two hyperfinite
Type II algebras, up to isomorphism. 
 A hyperfinite Type II$_1$ algebra has a trace defined for all elements of the algebra; its representations are labeled by a positive real number $x$ (it is also
 possible to have a representation with $x=\infty$; this happens if the commutant of the Type II$_1$ algebra is of Type II$_\infty$).
A hyperfinite Type II$_\infty$ algebra has a trace that is defined only for some elements of the algebra; when it acts on a Hilbert space,
its commutant is of Type II$_1$ or Type II$_\infty$, and thus it has  distinct kinds of Hilbert space representation. (A representation in which the commutant is
of Type II$_\infty$ is unique up to isomorphism; a representation in which the commutant is of Type II$_1$ can be classified by a positive real number $x$,
but there is no canonical way to normalize $x$.)

(3) A Type III algebra $\A$ has no irreducible representation in a Hilbert space; its commutant in any representation  
is another algebra isomorphic to $\A$.
A Type III algebra has no trace, and its representations are all isomorphic.  In the thermodynamic limit, 
the natural algebra of observables of a generic system in quantum statistical mechanics
 is of Type III$_1$.    The other hyperfinite Type III algebras are the Type III$_\lambda$ algebras that we have
 described, and a Type III$_0$ algebra that is a kind
 of $\lambda\to0$ limit of a Type III$_\lambda$ algebra.

\subsection{Relation To Quantum Field Theory}\label{reln}

In section \ref{QFT}, in order to understand quantum field theory in a curved spacetime $M$, especially in the case of an open universe, it was important
to consider the algebra $\A_\U$ of observables in a local region $\U\subset M$.    Now that we have some experience with von Neumann algebras,
we can return to this subject and ask what kind of algebra is $\A_\U$.     

We consider again the setup discussed in section \ref{whatis}.    We take $S$ to be a Cauchy hypersurface in $M$, which
we assume compact for simplicity, and $B$ to be a codimension 1 submanifold of $S$   whose
complement  consists of two open sets $V$ and $V'$ (fig. \ref{Bounded}).   We let $\U$ and $\U'$ be the domains of dependence of $V$ and $V'$ in $M$.
Then $\A_\U$ and $\A_{\U'}$ are the algebras
of observables in bounded regions of spacetime.   

To identify the algebra $\A_\U$, we use the simple remarks about von Neumann algebras that were explained at the end of section \ref{lss}.   
We also use a few facts about quantum field theory.   One set of facts involves entanglement.  For any state $\Psi\in\H$, there is a universal divergence in the entanglement entropy between
modes in $\U$ and modes in $\U'$.   The divergence comes from short wavelength modes that are localized near $B$, the ``entangling surface'' that separates
$\U$ and $\U'$.   The leading divergence is universal in the sense that it does not depend on the choice of $\Psi$.   The reason for this is that the divergence
comes from modes of very short wavelength, and at short distances every state in $\H$ looks the same.  Because it comes from modes localized near $B$,
the leading divergence is proportional to the area of $B$.   The divergence involves pairs of modes on opposite sides of $B$ that are entangled. The geometry of $\U$ and $\U'$ near $B$ can
be modeled locally by two opposite Rindler wedges in Minkowski space (see fig. \ref{Bounded}(b)), so the entanglement structure at short distances is given by Unruh's thermal
interpretation of Rindler space \cite{Unruh}, or equivalently by the Bisognano-Wichman theorem \cite{BW} which describes the application of Tomita-Takesaki
theory to Rindler space.\footnote{The contributions of Unruh and of Bisognano and Wichman were done independently at about the same
time.   The connection between them was appreciated only years afterward, by Sewell \cite{Sewell}.}  The thermal description shows that many different
pairs of modes have all possible degrees of entanglement.   In fact, the modular Hamiltonian of Rindler space is a Lorentz boost generator,
which has a continuous spectrum consisting of all real numbers.  

We also need the fact  that the local algebras in quantum field theory do not have a trace.  Given any $\a\in\A_\U$,  
its trace in $\H$ is divergent because of the infinity
of things that could be happening in region $\U'$, and there is no regularized version of this trace.

Given these facts, we can guess what type of algebra is $\A_\U$.   The algebras whose structure incorporates a universal divergence in the entanglement
entropy are of Type II and Type III.   The absence of a trace tells us that $\A_\U$ is of Type III.   And the fairly generic pattern of entanglement, with many 
 modes of all possible degrees of entanglement, analogous
to a rather generic choice of $E_n$'s in eqn. (\ref{nextcasse}),  indicates that the algebra is of Type III$_1$.

This is believed to be the right answer.  Rigorous arguments are available  \cite{Fred}, at least for open sets in Minkowski space.  Informal summaries
of these arguments with somewhat more detail than provided here can be found
 in \cite{HaagBook}, section V.6 and in \cite{WittenNotes}, section 6.5. 
 
In either quantum field theory or statistical mechanics, the Type III nature of the algebras is associated, as we have seen, 
to a divergence in the entanglement entropy.
In quantum field theory, this is an ultraviolet divergence between modes inside and outside the entangling surface $B$; in statistical mechanics
it is an infrared divergence between the modes of the left and right systems.   Even though one divergence 
is an ultraviolet divergence and the other
is an infrared divergence, they are actually quite similar.  

 \begin{figure}
 \begin{center}
   \includegraphics[width=4.5in]{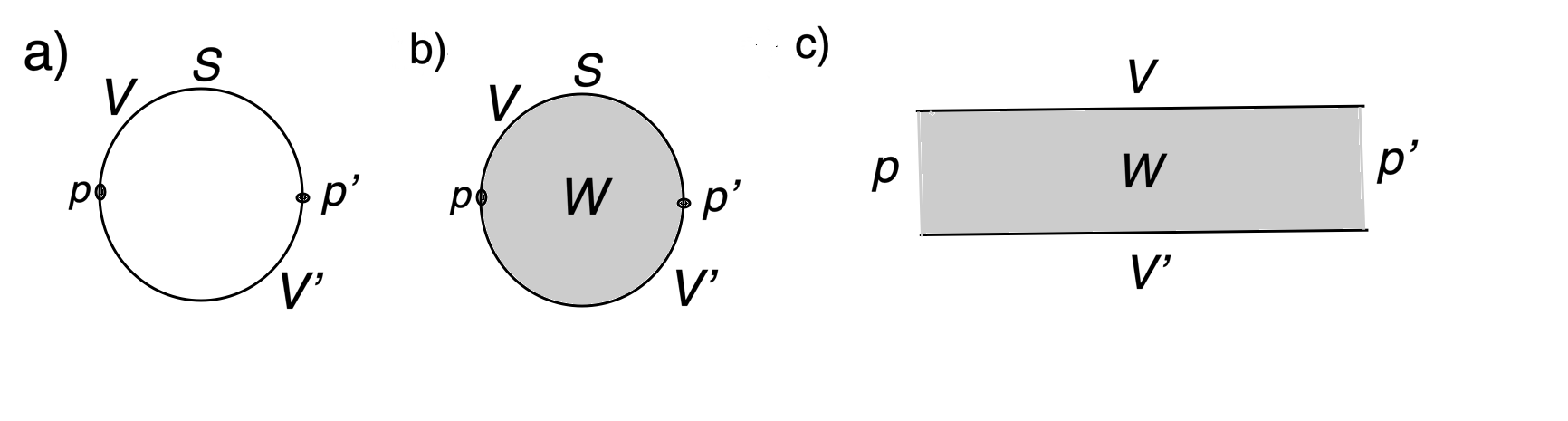}
 \end{center}
\caption{\small  (a) In a two-dimensional spacetime, we consider
a  Cauchy hypersurface consisting of a circle $S$.  The entangling surface $B$
is a pair of points $p,p'$, whose complement in $S$ consists of the two intervals $V$ and $V'$.  (b) A path integral
on the disc $W$, whose boundary is $S$, computes the ground state of a conformal field theory.
(c)  The disc with the points $p$ and $p'$ removed can be conformally mapped to
$\R\times I$, where $I$ is a unit interval.  $V$ and $V'$ become the two boundary 
components of $\R\times I$, and the points $p,p'$ are projected to infinity.  \label{Conformal}}
\end{figure} 

 One aspect of the similarity is that in the case of a conformal field theory (CFT),
one can actually make a conformal transformation between the two types of divergence.
This is best-known in the case of a two-dimensional CFT \cite{HLW,CC}.  One takes the initial value surface $S$ to be a 
circle, and one takes the entangling surface $B$
to consist of a pair of points $p,p'\in S$; the complement of these points is the union of two open intervals 
$V$ and $V'$ (fig. \ref{Conformal}(a)).   Let $\H$ be the Hilbert space of the given theory
formulated on $S$.   The conformally invariant
ground state $\Omega\in\H$ can be computed by a path integral on a disc $W$ whose boundary is $S$ (fig. \ref{Conformal}(b)).  The state $\Omega$
is highly entangled between regions $V$ and $V'$.   If one tries to quantify this entanglement by computing the entanglement entropy
between $V$ and $V'$, one runs into
an ultraviolet divergence which is related to the fact that the local algebras actually are of Type III.   This ultraviolet 
divergence can be converted into
an infrared divergence by a conformal mapping that maps the disc $W$, minus the two points $p$ and $p'$, to an infinite 
strip $\R\times I$, where $I$ is an interval (fig. \ref{Conformal}(c)).    
After a Weyl
rescaling, one can take the metric of $\R\times I$ to be the flat metric $\d x^2+\d\tau^2$, $-\infty<x<\infty$, $0\leq\tau \leq 1$.
The path integral on this strip describes the thermofield double of the theory under study on the noncompact initial value 
surface $S'=\R$, and with inverse temperature 1;
the two open intervals $V$ and $V'$ have become the boundaries of the strip at $\tau=0,1$, corresponding to 
the two copies in the thermofield double.
The points $p$ and $p'$ have been projected to $x=\pm \infty$.    In the description on the disc, there is an 
ultraviolet divergence in the entanglement entropy of the
vacuum that arises from the behavior
of modes on opposite sides of the points $p$ or $p'$.     In the description on the strip, this divergence is 
reinterpreted as an infrared divergence  in the entanglement entropy of the thermofield
double near $x=\pm \infty$.

All this has an analog for a conformally invariant theory in any dimension $D$  \cite{Cas,Cas2}.   
One takes the initial value surface $S$ to be a $(D-1)$-sphere with a round
metric.  One further takes the entangling surface $B$ to be an ``equator'' in $S$, and $V$ and $V'$ to be 
the northern and southern hemispheres. Thus the setup is the same as before, except that $S$ is a $(D-1)$-sphere instead of
a circle, and for $D>2$, the equator in $S$ is connected, rather than consisting of two isolated points.    The ground state
$\Omega$ of a $D$-dimensional CFT
 can be computed by a path integral on a flat unit ball $W$ in $D$ dimensions,  whose boundary is $S$.   
 After removing the equator $B$ of the boundary, $W$
 can be conformally mapped to a product $H\times I$, where $H$ is hyperbolic space of dimension $D-1$, and 
 $I$ is again the interval $0\leq \tau\leq 1$.
A path integral on $H\times I$ again describes a thermofield double at termperature 1.    The only 
difference from before is that the spatial
manifold $H$ on which the thermodynamics is defined is hyperbolic space of dimension $D-1$ rather than 
Euclidean space.   But $H$ has infinite volume just like Euclidean space, and
there is an infrared divergence in the entanglement entropy between the two sides of the thermofield double.   
This divergence matches the ultraviolet divergence
between modes on $V$ and modes on $V'$ in the original description on the sphere.

\subsection{The Hagedorn Temperature}\label{ht}

In this section, we return to the Araki-Woods model of a Hamiltonian that is a sum of single-qubit Hamiltonians $H=\sum_{n=1}^\infty H_n$, with
\be\label{believe} H_n=\begin{pmatrix} 0 & 0\cr 0 & E_n \end{pmatrix}.\ee
Now, however, we assume that $\lim_{n\to\infty} E_n=+\infty$.   We also assume for convenience that the $E_n$ are all positive, so that the
ground state of the $n^{th}$ qubit is the state $|\negthinspace\uparrow\ra_n=\begin{pmatrix}1\cr 0\end{pmatrix}$.   The ground state of the whole system
is $|\Omega\ra_\uparrow=\otimes_{n=1}^\infty |\negthinspace\uparrow\ra_n$.    As in section \ref{spins}, a Hilbert space $\H^*$ that describes all states of infinitely many qubits
is nonseparable.   But we can define a separable Hilbert space $\H$ in which
every state can be approximated by states in which all but finitely many qubits are in their ground state $|\negthinspace\uparrow\ra_n$.   

$\H$  has a basis consisting of states in which all qubits except some finite subset $n_1,n_2,\cdots, n_k$ are in their ground state.
The energy of such a state is $E=E_{n_1}+E_{n_2}+\cdots + E_{n_k}$.   Now let us ask this question:   What is the probability to observe such a state
in a thermal ensemble, at inverse temperature $\beta$?   The probability to observe
 a microstate of energy $E$ is  of course $p(E,\beta)=\frac{1}{Z(\beta)}e^{-\beta E}$, where
\be\label{tofo} Z(\beta) =\prod_{n=1}^\infty (1+e^{-\beta E_n})\ee
is the partition function.   In section \ref{td}, we studied models in which the partition function diverges in the limit of infinitely many qubits, but now we wish
to consider the case that the $E_n$ tend to infinity quickly enough that $Z(\beta)$ converges, at least for some range of $\beta$.
Clearly when $Z(\beta)<\infty$, we have $p(E,\beta)>0$ and the probability to observe any given microstate in $\H$ in the thermal ensemble is positive.

This suggests that the thermal ensemble at inverse temperature $\beta$ can simply be defined in the Hilbert space $\H$.   To see that this is the
case, let us ask what is the probability that, for some $N$,  all qubits except the first $N$ are in their ground state.   The probability for this
in the thermal ensemble is $Z_N(\beta)/Z(\beta)$, where $Z_N(\beta)$ is a truncated partition function defined for the first $N$ qubits:
\be\label{tofoo} Z_N(\beta) =\prod_{n=1}^N (1+e^{-\beta E_n}).\ee
If $Z(\beta)=\infty$, then $Z_N(\beta)/Z(\beta)=0$ for all $N$, since $Z_N(\beta)$ is finite, but if $Z(\beta)<\infty$, then
\be\label{zofo}\lim_{N\to\infty}\frac{Z_N(\beta)}{Z(\beta)}=1. \ee
For $Z(\beta)<\infty$, this shows that, with probability 1, any state in the thermal ensemble can be approximated arbitrarily well by states in $\H$.

Therefore, in this situation, it is not necessary to go to the thermofield double in order to describe the thermal ensemble of an infinite number of qubits
in a separable Hilbert space.   What happens if we do go to the thermofield double anyway?   The answer is that we get nothing essentially new; when $Z<\infty$
the thermofield double does not combine the two copies in an interesting way.

To explain how this comes about, we consider two copies of this system,\footnote{The system considered has an obvious time-reversal symmetry, so the
two copies are identical.} the left and right copy, each consisting of countably many qubits with identical Hamiltonians.   There are nonseparable
Hilbert spaces $\H_\ell^*$ and $\H_r^*$ that describe all possible states of the left and right systems, and these  contain separable subspaces $\H_\ell$ and $\H_r$
that have bases with finitely many qubits excited.  We also write $\A_{\ell,0}$ or $\A_{r,0}$ for the algebras of operators that act on finitely many qubits
of the left or right systems, respectively, and $\Omega_{\uparrow,r}$, $\Omega_{\uparrow,\ell}$ for the ground states of the right and left systems.

We can define the thermofield double state in the naive way as a vector in $\H_\ell^*\otimes \H_r^*$:
\be\label{nozacko} \Psi_\TFD=\frac{1}{\sqrt {Z(\beta)}}\bigotimes_{n=1}^\infty \begin{pmatrix}1& 0\cr 0& e^{-\beta E_n/2}\end{pmatrix}.\ee
Then we define the thermofield double Hilbert space $\H_\TFD$ as the closure of the set of all vectors of the form $\a_r\Psi_\TFD$, $\a_r\in\A_{r,0}$.

In section \ref{lss}, we could not define the thermofield double state and Hilbert space of an infinite system in this simple way, because we were considering systems
with $Z(\beta)=\infty$.   Therefore, in eqns. (\ref{zacko}) and (\ref{ozacko}), we defined the thermofield double state directly by a formula analogous to (\ref{nozacko})
for a finite system of $N$ qubit pairs, and  we invoked the GNS construction to define $\Psi_\TFD$ and $\H_\TFD$ in the limit of an infinite system.

{\it A priori}, eqn. (\ref{nozacko})  defines $\Psi_\TFD$ as a vector in the nonseparable Hilbert space $\H_\ell^*\otimes \H_r^*$, and likewise it defines $\H_\TFD$
as a subspace of that space.   But actually, when $Z(\beta)<\infty$ so that the definition (\ref{nozacko}) makes sense, $\H_\TFD$
is the same as the separable Hilbert space $\H_\ell\otimes\H_r$.  To see this, observe that a truncated version of $\Psi_\TFD$, namely
\be\label{noozacko} \Psi_{\TFD,N}=\frac{1}{\sqrt {Z(\beta)}}\bigotimes_{n=1}^N \begin{pmatrix}1& 0\cr 0& e^{-\beta E_n/2}\end{pmatrix}\bigotimes_{n=N+1}^\infty
\begin{pmatrix} 1& 0 \cr 0 & 0 \end{pmatrix}\ee
is manifestly contained in $\H_\ell\otimes \H_r$.    But $\lim_{N\to\infty}\Psi_{\TFD,N}=\Psi_{\TFD}$.   So $\Psi_{\TFD}\subset \H_\ell\otimes \H_r$.
This implies that $\H_\TFD\subset \H_\ell\otimes \H_r$, since $\H_\TFD$ is generated from $\Psi_\TFD$ by the action of $\A_{r,0}$, and $\H_\ell\otimes \H_r$
is closed under the action of $\A_{r,0}$.
Conversely,  $\A_{r,0}$ contains an operator $\Pi_N$ that projects onto the ground state of the first $N$ qubits, and acts trivially on others;
since $|\Omega\ra_{\uparrow,\ell}\otimes |\Omega\ra_{\uparrow,r}=Z(\beta)\lim_{N\to\infty}\Pi_N\Psi_\TFD$, we have $|\Omega\ra_{\uparrow,\ell}\otimes |\Omega\ra_{\uparrow,r}\in \H_\TFD$, implying, since $\H_\ell\otimes\H_r$ is generated from $|\Omega\ra_{\uparrow,\ell}\otimes |\Omega\ra_{\uparrow,r}$
by acting with $\A_{\ell,0}\otimes \A_{r,0}$, that $\H_\ell\otimes \H_r\subset \H_\TFD$.

In short, when $Z(\beta)<\infty$, the thermofield double Hilbert space $\H_\TFD$ is just a simple tensor product $\H_\TFD=\H_\ell\otimes \H_r$.
Likewise, if we take the closures of $\A_{\ell,0}$ and $\A_{r,0}$ to get von Neumann algebras $\A_\ell$ and $\A_r$  that act on 
the thermofield double, we get nothing essentially new:
$\A_\ell$ and $\A_r$ are simply the Type I$_\infty$ algebras of all bounded operators on $\H_\ell$ and $\H_r$, respectively. In this situation,
the Hamiltonians $H_\ell$ and $H_r$ of the left and right systems are well-defined as operators on $\H_\TFD$ and are elements of $\A_\ell$ and $\A_r$
respectively.    All in all, when $Z(\beta)<\infty$, we gain little by going to the thermofield double; it is just a tensor product of two decoupled systems.

On the other hand, if $Z(\beta)=\infty$, we are in the situation described in section \ref{lss}.   The thermofield double Hilbert space $\H_\TFD$ is an essentially
new construction, not a simple tensor product, and the corresponding algebras $\A_\ell$ and $\A_r$ are generically of Type III$_1$.   The operators $H_\ell$
and $H_r$ cannot be defined as operators on $\H_\TFD$.   As a partial substitute one has $\h H=H_r-H_\ell$, which does have a limit for the infinite system,
and generates a group of outer automorphisms of $\A_r$ and of $\A_\ell$.

A particularly interesting case, and important background for section \ref{large}, is the borderline case in which $Z(\beta)<\infty$ at low enough temperature, but $Z(\beta)=\infty$ at sufficiently high temperature.
This can happen if $E_n\sim c\log n$ for large $n$, with a constant $c$.   Then at low temperatures, the thermofield double construction just describes two
decoupled systems, and all algebras are of Type I.   But at high termperatures, the thermofield double construction is genuinely new, combines the two systems in a subtle way,  and is essential
if we wish to describe the infinite qubit system in a separable Hilbert space.   The algebras are of Type III, and time translations of the algebras
are a group of outer automorphisms.    By analogy with standard terminology in strong interaction theory, string theory, and large $N$ gauge theory,
we call the temperature at which $Z(\beta)$ ceases to converge the Hagedorn temperature $T_H$, and we also write $\beta_H=1/T_H$.

\subsection{Density Matrices and Entropy}\label{density}

For a quantum system with a Hilbert space $\H$ of physical states, one usually defines the von Neumann entropy, for a state of the system described
by a density matrix $\rho:\H\to\H$, by $S(\rho)=-\Tr\,\rho\log \rho$.  As usual, here $\rho$ is a positive self-adjoint operator normalized to $\Tr\,\rho=1$.
The trace function $\a\to \Tr\,\a$  is a complex linear function from the algebra $\A$ of all operators
on $\H$ (or all bounded operators if $\H$ is infinite-dimensional) to $\C$; it satisfies $\Tr\,\a^\dagger\a>0$ for all $\a\not=0$.   This last condition ensures
that the function $F(\a)=\Tr\,\a\rho$ satisfies the conditions as stated in section \ref{whatis} for a state on the algebra $\A$. 
Conversely, every state on the algebra $\A$ is of the form $F(\a)=\Tr\,\a\rho$ for some density matrix $\rho$.  To prove this, one first observes that
the trace is nondegenerate, in the sense that every linear function on $\A$ is of the form $F(\a)=\Tr\,\a\b$ for some $\b\in\A$.   The other axioms
for a state (namely $F(\a^\dagger\a)\geq 0$ and $F(1)=1$) imply that $\rho$ must be a positive operator satisfying $\Tr\,\rho=1$, in other words, a density matrix.

Now consider any von Neumann algebra $\A$ that acts on a Hilbert space $\H$.   An element of $\A$ is called positive if it corresponds to a positive,
self-adjoint operator on $\H$.    If the algebra $\A$ has a nondegenerate trace,  
then the same arguments as before show that any state on $\A$
is of the form $F(\a)=\Tr\,\a\rho$, where $\rho$ is a density matrix, that is, a positive element of $\A$ of trace 1.   We can then define, for any state,
the von Neumann entropy $S(\rho)=-\Tr\rho\log\rho$.   In this article, the von Neumann algebras that we consider are factors, that is, algebras with trivial
center (in other words, algebras whose center consists only of $\C$).   A factor is the von Neumann algebra analog of a simple Lie group.   
In the case of a factor, any trace is nondegenerate
and therefore the condition to be able to define a von Neumann entropy is only that $\A$ has a trace.   Also, in the case of a factor, if a trace exists,
it is unique up to a multiplicative constant.   As we will see, in general it is important that the trace is not quite unique.

Let us discuss entropy for the different types of von Neumann algebra.   The original arena for von Neumann entropy
is a Type I algebra, which acts irreducibly on
a Hilbert space $\H$.    A standard argument shows that $S(\rho)$ is always nonnegative in this case.   A subtlety is that in the case of a Type I$_\infty$ algebra,
that is, if $\H$ is infinite-dimensional, it is possible to have $S(\rho)=+\infty$.   That is because the condition $\Tr\,\rho=1$ does not ensure the convergence
of $\Tr\,\rho\log\rho$.  Thus for Type I$_\infty$, the entropy takes values in $\R_{\geq 0}\cup+\infty$.   

Now let us consider algebras of Type II and Type III.   A Type III algebra has no trace and therefore no notion of von Neumann entropy.   However, a Type II algebra does have a trace, as explained in section \ref{lss}, and therefore one can define an entropy\footnote{This has been discussed, primarily for Type II$_1$, in
\cite{IESegal,ConnesStormer}.}  for a state of a von Neumann
algebra of Type II.

Let us explore this notion, initially for the case of Type II$_1$.   We recall that a Type II$_1$ algebra $\A$ has a trace
that is defined for all elements of $\A$, and is conventionally normalized so that $\Tr\,1=1$.   The simplest density matrix to consider is $\rho=1$.
In this case, $\log\rho=0$ so $S(\rho)=-\Tr\,\rho\log\rho=0$.   But what $\rho$ describes is far from being a zero entropy state in any conventional sense.   In fact, the density matrix
$\rho=1$ describes the maximally entangled thermofield double state $\Psi_\TFD$ that was defined in eqn.  \ref{zacko}.  Indeed, we defined the trace in a Type II$_1$
algebra by $\Tr\,\a=\la\Psi_\TFD|\a|\Psi_\TFD\ra$.   This implies that to satisfy $\Tr\,\a\rho = \la\Psi_\TFD|\a|\Psi_\TFD\ra$, we need to take $\rho=1$,
so $\rho=1$ is the density matrix that corresponds to the maximally entangled thermofield double state $\Psi_\TFD$.

The maximally entangled state of $N$ qubit pairs has von Neumann entropy $N\log 2$, and to get a Type II$_1$ algebra, one takes the limit $N\to\infty$.   
So the entropy diverges for $N\to\infty$. But 
with the Type II$_1$ definition of entropy, the large $N$ limit of the thermofield double 
state is deemed to have entropy 0.   This suggests that in general Type II$_1$ entropy
is ordinary von Neumann entropy with the entropy of a maximally entangled state subtracted. 
To confirm this interpretation, recall that a  
general state of a Type II$_1$ algebra can be associated  to a system of $N$ qubit pairs, in the large $N$ limit, in a state in which  
 almost all the qubit pairs are almost maximally entangled.   
 As an illustrative example, suppose that the first $k$ qubits are definitely in a ``spin up'' state, and the others are maximally mixed.
 To describe this state by a density matrix,
   let $\Pi$ be the orthogonal  projection operator onto the subspace
of  states of $N$ qubits in which the first $k$ qubits  all have spin up.   For finite $N$,  $\rho= 2^{-(N-k)}\Pi$ is a density matrix of the $N$ qubit system, and
its von Neumann  entropy is  $S_N=(N-k)\log 2$.   But in  the limit $N\to\infty$, $\Pi$ becomes an element    of the Type II$_1$ algebra with
$\Tr\,\Pi=2^{-k}$.   So in the Type II$_1$ algebra, we can define a density matrix $\rho=2^k\Pi$, with $\Tr\,\rho=1$.  For this density matrix, we compute $S=-\Tr\,\rho\log\rho=-k\log 2$.
This is the same as $S=\lim_{N\to\infty}(S_N-N\log 2)$.   In other words, $S$ is the large $N$ limit of $S_N-N\log 2$, or differently put, it is the large $N$ limit of
the entropy difference between the actual state of the first $N$ qubits and a maximally mixed state.
  As this example and discussion suggest, entropy is negative definite for a Type II$_1$ factor.

A Type III algebra can similarly be constructed, as described in section \ref{lss}, from an infinite collection of (non-maximally) entangled qubit pairs.   But in this
case, there is no way, using only the structure of the algebra, to define a renormalized entropy.   This is reflected in the fact that the divergences in the entanglement
entropy of a local region in quantum field theory are model-dependent (there is a leading divergence proportional to the area of the boundary of the
region, with a theory-dependent coefficient, and there are subleading divergences that depend on the spacetime dimension and the operator content of the theory).  
Renormalized entropies of local regions in quantum field theory can be useful and important, but defining them requires information beyond the von Neumann algebra
structure.  Similar remarks apply in quantum statistical mechanics if one considers the entropy of a state that is not thermal but looks thermal near spatial infinity.

Now let us consider an algebra $\A$ of Type II$_\infty$.   Such an algebra can be factored as\footnote{Because of the Type I$_\infty$ factor, entropy
in a Type II$_\infty$ factor can take the value $+\infty$.   Because of the Type II$_1$  factor (or because of the additive indeterminacy that we discuss shortly)
it can be negative.   In fact, it can take any value from $-\infty$ to $+\infty$.}  $\A=\A_1\otimes \A_2$, where $\A_1$ is of Type II$_1$
and $\A_2$ is the  type I$_\infty$ algebra of bounded operators on a separable Hilbert space $\K$.   We can define a trace $\Tr_1$ on $\A_1$ normalized so that $\Tr_1\,1=1$, and a trace $\Tr_2$ on $\A_2$ normalized so that,
if $\Pi\in \A_2$ is the orthogonal projection operator on a one-dimensional subspace $\K_0\subset \K$,
then $\Tr_2\,\Pi=1$.   Then we can define a trace on $\A$ by $\Tr=\Tr_1\otimes \Tr_2$.

Since a trace is available, there is a notion of entropy for such an algebra $\A$.    But there is a very important subtlety: there is no canonical way to
normalize this trace.   One immediate observation is that, since the identity element of $\A_2$  has a divergent trace, we cannot 
normalize the trace, as we did for Type II$_1$, by saying that the identity element
has trace 1.   A deeper explanation involves the fact that a Type II$_\infty$  algebra can be factored as the tensor product of algebras of Type II$_1$ and Type
I$_\infty$ in many different ways.

 Let
$\A_3$ be an algebra of Type I$_n$, acting on an $n$-dimensional Hilbert space $\H_n$, and define the trace $\Tr_3$ on $\A_3$ to be the usual trace of
a linear transformation acting on $\H_n$. Thus $\Tr_3\,1=n$.   Consider the algebra $\A=\A_1\otimes\A_2\otimes \A_3$, with $\A_1$ and $\A_2$ as before.
The algebra $\A_1'=\A_1\otimes \A_3$ is of Type II$_1$, and the algebra $\A_2'=\A_2\otimes \A_3$ is of Type I$_\infty$.   So $\A=\A_1'\otimes \A_2
=\A_1\otimes \A_2'$ gives two different factorizations of the same Type II$_\infty$ factor as the tensor product of factors of Type II$_1$ and Type I$_\infty$.
These two factorizations of $\A$ lead to definitions of the trace that differ by a factor of $n$.  To see this, consider the element $\a=1\otimes \Pi\otimes 1\in
\A_1\otimes \A_2\otimes \A_3$, where as before $\Pi$ is an element of $\A_2$ that projects on a one-dimensional subspace $\K_0\subset \K$.  With respect to the factorization of $\A$ as $\A_1'\otimes \A_2$, $\a$ is the tensor product of $1\in \A_1'$ with $\Pi\in \A_2$,
so the normalization of the trace that is natural for this factorization gives $\Tr\,\a=\Tr\,1\otimes \Pi=1$.
On the other hand, with respect to the factorization of  $\A$ as $\A_1\otimes \A_2'$, $\a$ is $1\otimes \Pi'$, where $\Pi'$ projects on the $n$-dimensional subspace
$\K_0\otimes \H_n\subset \K\otimes \H_n$.    So with the normalization of the trace that is natural for this second factorization, we would have $\Tr\,\a=n$.

Thus different choices of factorization motivate different normalizations of the trace for a Type II$_\infty$ factor.
In fact, there can be no natural normalization of the trace, because 
 a Type II$_\infty$ algebra has an outer automorphism group that rescales the trace by any positive
real number.
(This is a standard result in von Neumann algebra theory and will be explained elsewhere \cite{WittenNew}.) 
However, as long as we assume our algebra to be a factor, analogous to a simple Lie group, the only indeterminacy in the trace is an overall
multiplicative constant.

Let us see how this indeterminacy affects the definition of entropy.   If we rescale the trace by $\Tr\to\lambda\Tr$, we must compensate by rescaling the density
matrix by $\rho\to\lambda^{-1}\rho$, to preserve $\Tr\,\rho=1$.   Under the combined rescaling of $\Tr$ and $\rho$, we have
\be\label{wonko}S(\rho)=-\Tr\,\rho\log\rho\to S(\rho)+\log \lambda.\ee
In other words, when we rescale the trace by a factor $\lambda$, the entropy of any state is shifted by an additive constant $\log\lambda$, independent of the state.

This means that entropy in a Type II$_\infty$ factor is somewhat analogous to entropy in classical mechanics.   The concept of entropy was originally discovered
in the 19th century by macroscopic arguments, which showed that there must be a state function $S$ of a system in thermal equilibrium that obeys the first
law of thermodynamics.  If the relevant thermodynamic variables are the energy $E$, temperature $T$, pressure $p$, and volume $V$, then the first
law reads
\be\label{tofu}\d E=T\d S-p\d V.\ee
This equation determines $S$ up to an additive constant.   Quantum mechanically, one can fix the additive constant by saying that (assuming the system
under study has a unique ground state) the entropy vanishes at $T=0$.  Classically, no such statement is possible, because for example the entropy of a classical
harmonic oscillator goes to $-\infty$ for $T\to 0$.   

For another explanation of why it is 
difficult in classical physics to fix an overall additive constant in the entropy, consider a system of particles with positions $\vec x$ and momenta $\vec p$.
One can describe one's state of knowledge about the system with a probability distribution function $\rho(\vec p, \vec x)$.   Then a 
classical definition of entropy is
\be\label{defent} S=-\int\d\vec p\,\d \vec x \rho(\vec p,\vec x)\log \rho(\vec p,\vec x). \ee
The integral runs over the classical phase space of the system.   But classically, there is no natural way to normalize the phase space measure, since, for any
pair of canonical variables $p,x$, the measure $\d p\,\d x$ has dimensions of action, and classically there is no natural constant with dimensions of action that could
be used to normalize it.
(Quantum mechanically, one replaces $\d p\,\d x$ by $\d p\,\d x/2\pi\hbar$.)    The phase space measure in classical mechanics is naturally defined up to an overall
multiplicative constant, but there is no natural way to fix this constant.   If we rescale the measure by $\d \vec p\,\d\vec x\to \lambda\d\vec p\,\d\vec x$, for a positive
constant $\lambda$, then we have to compensate by $\rho\to\lambda^{-1}\rho$.  But this then shifts $S$ by $S\to S+\log\lambda$, just as in the discussion of
a Type II$_\infty$ algebra.   

Here is an interesting example of entropy for a state of a Type II$_\infty$ algebra $\A$. Pick a factorization $\A=\A_1\otimes \A_2$ where $\A_1$ is of Type II$_1$ and $\A_2$ is of
Type I$_\infty$.   Let $\Pi_k,$ $k=1,\cdots, n$, be orthogonal projection operators in $\A_1$ satisfying $\Pi_i \Pi_j=\delta_{ij}\Pi_j$, $\Tr\,\Pi_i=p_i$,
$\sum_{k=1}^n \Pi_k=1$.   Let $\rho_k$, $k=1,\cdots, n$  be any density matrices in $\A_2$.   Then $\rho=\sum_{k=1}^n \Pi_k \otimes \rho_k$  is a density
matrix in $\A$.   Its entropy is
\be\label{entis}S(\rho)=\sum_k p_k S(\rho_k) =-\sum_k p_k \Tr\, \rho_k\log \rho_k. \ee
In verifying this, one uses the fact that $\Pi_k\log \Pi_k=0$ for all $k$, since the only eigenvalues of $\Pi_k$ are 0 and 1, and $x\log x=0$ if $x=0$ or 1.
It is noteworthy that in the formula for $S(\rho)$ there is no Shannon term $-\sum_k p_k\log p_k$.   Thus in this particular example, Type II$_\infty$ entropy
is the average entropy of an ensemble of density matrices.

We have phrased this discussion in terms of the von Neumann entropy, but similar remarks apply for other information theoretic
measures such as the Renyi entropies $S_\alpha(\rho)=\frac{1}{1-\alpha}\log \Tr\,\rho^\alpha$.  These entropies can be defined for a state on an algebra of
Type I or Type II (but not, of course, Type III).   In the Type II$_\infty$ case, if we rescale the trace by a factor $\lambda$ and compensate by rescaling the
density matrix $\rho\to\lambda^{-1}\rho$, the Renyi entropies are shifted by
\be\label{todlo} S_\alpha(\rho)\to S_\alpha(\rho)+\frac{\alpha}{\alpha-1} \log\lambda.\ee
Thus all the Renyi entropies are well-defined in Type II$_\infty$, up to a shift that is controlled in 
this way by the same $\alpha$-independent parameter $\lambda$ that controls the indeterminacy in the von Neumann entropy.

 \section{The Large $N$ Limit and the Thermofield Double}\label{large}
 
QCD in Minkowski space is confining at low temperatures and has a deconfinement transition at a certain positive temperature \cite{CabPar,Poly,Suss}.
In finite volume, this transition is smoothed out.   But if we replace QCD, which has gauge group $SU(3)$, with a theory with gauge group $SU(N)$
(or $SO(N)$ or $Sp(N)$), then it is believed  that  in the large $N$ limit, there is a sharp deconfinement transition even in finite volume.   This is true
even for zero \cite{Other,HPe,Poly2,Aha} or small \cite{AhaTwo} coupling.  In the case of a gauge theory that has a gravitational dual, this  phenomenon is relevant to the AdS/CFT correspondence.   In that application, if the bulk
spacetime has $D$ noncompact dimensions, so that the conformal boundary has dimension $D-1$, the spatial manifold on which the gauge theory lives is most simply taken to be a sphere
$\S^{D-2}$.   Then the deconfinement transition is dual to the Hawking-Page transition \cite{HP} between a thermal gas in Anti de Sitter
space and a black hole \cite{Wit}.   We call the temperature at which this transition occurs $T_\HP$.

Let $\H_N$ be the Hilbert space of  a large $N$ gauge theory, quantized on the compact manifold 
$\S^{D-2}$. It is believed that  $\H_N$ has a large $N$ limit $\H_\infty$: the energy eigenvalues and multiplicities  and the matrix elements of operators
all have limits for $N\to\infty$ (and in particular the multiplicities do not grow with $N$).
At sufficiently low temperatures, the thermal ensemble can be understood in the limiting Hilbert space $\H_\infty$.   Because the energy eigenvalues
and multiplicities have limits for $N\to\infty$, the thermodynamic functions such as entropy, energy, etc., all have large $N$ limits at sufficiently
low temperatures.    Thus in the low temperature phase, the energy and entropy are of order $N^0$ for large $N$.
In the range of temperatures at which the thermodynamics can be described in $\H_\infty$, 
the natural algebra of observables at large $N$ is simply the Type I$_\infty$ algebra of all operators on $\H_\infty$.

It is believed that in large $N$ gauge theories,
 the partition function $\Tr_{\H_\infty}\,e^{-\beta H}$, computed in the limiting large $N$ Hilbert space $\H_\infty$, diverges at a temperature
$T_H$, known as the Hagedorn temperature.
This can be seen quite explicitly at zero coupling.  At or below $T_H$, the description based on the Hilbert space
$\H_\infty$ has to break down.   We will loosely call the transition  at which this happens the Hagedorn
transition.   In various weakly coupled gauge theories, it is believed that the transition occurs at a temperature slightly below $T_H$ \cite{AhaTwo}.   
In gauge theories -- such as $\N=4$ super Yang-Mills theory in four dimensions -- that participate in AdS/CFT duality, the transition away from a description in 
the Hilbert space $\H_\infty$ occurs
at the Hawking-Page temperature $T_\HP$, which is far below $T_H$ (assuming that $g^2N$ is large and the AdS theory can be studied semiclassically).  In fact,
   $T_H$ is at the string scale, while $T_\HP$ is of
order the inverse radius of the AdS space.   
Thus in the AdS/CFT context, what we are calling the Hagedorn transition is the same as the Hawking-Page
transition and occurs at a temperature far below $T_H$.   Regardless, above this transition the physics can no longer be described in $\H_\infty$.

Above the transition temperature,
the energy and entropy are both proportional to $N^2$ for large $N$ \cite{Thorn} and in particular do not have large $N$ limits.
To reach the high temperature phase in the large $N$ limit, we have to take $N\to\infty$ with the energy  $E$ proportional to $N^2$
(as opposed to taking $N\to\infty$ with fixed $E$, which will always lead to a description in the Hilbert space $\H_\infty$, regardless of $E$).    In brief,
we will refer to the large $N$ limit in that regime as the large $N$ limit above the Hagedorn transition.
It seems very unlikely that there is a limiting large $N$ Hilbert space that describes the large $N$ limit above the Hagedorn transition.   
Certainly the literature does not contain any proposal for what such a large $N$ Hilbert space would be.
Since the entropy above the Hagedorn transition is of order $N^2$, when 
 one increases the rank of the gauge group from $N$ to $N+1$, the entropy increases by a multiple of $N$, and the Hilbert space becomes much bigger.   
 So most ``above the Hagedorn transition'' microstates with gauge group $SU(N+1)$ are ``new'' and have no antecedents in the $SU(N)$ theory.

The non-existence of a large $N$ limit of the Hilbert space above the Hagedorn transition would  be analogous to the non-existence, in quantum statistical
mechanics at temperature $T>0$, of a Hilbert space that describes the infinite volume limit.  In this analogy, $N^2$ plays
 the role of the volume $V$.  
There is also an analogy with the non-existence of
a natural Hilbert space description of quantum field theory in an open universe.   

The analogy with statistical mechanics also suggests a partial cure, which is to go to the thermofield double.   This is particularly powerful in gauge theories
that participate in AdS/CFT duality.   In such a case, the large $N$ limit of the thermofield double, in the high temperature phase, 
is a two-sided eternal black hole \cite{Malda}.   By quantizing the low energy bulk fields in the  black hole spacetime, one constructs
a Hilbert space $\H_\TFD$. This Hilbert space describes the large $N$ limit of the thermofield double of the gauge theory, in the high temperature phase.

A natural algebra of observables in the large $N$ limit is the algebra of polynomial functions of single trace operators. Based on the analogy with quantum
statistical mechanics and the discussion in section \ref{ht}, we can make a guess about the nature of this large $N$ algebra.   As already noted, below the Hagedorn
transition, the physics can be described in the original large $N$ Hilbert space $\H_\infty$, and the natural algebra acting on this Hilbert space is
of Type I.   In the high termperature phase, the analogy with statistical mechanics suggests that the natural algebra of single trace operators would be of
Type III.   Indeed, this has been argued recently by Leutheusser and Liu \cite{LL}, who made this claim partly based on the AdS/CFT duality, in  a paper with several novel ideas about quantum black holes.    These issues and some subtle corrections to the large $N$ limit
will be further discussed elsewhere \cite{WittenNew}.

Finally let us consider the Hamiltonians $H_r$ and $H_\ell$ that act on the two copies in the thermofield double.   
What happens in the large $N$ limit above the Hagedorn transition is similar
to what happens in the more familiar infinite volume limit.     The difference $\h H=H_r-H_\ell$ annihilates the thermofield double state $\Psi_\TFD$ and
converges, in the large $N$ limit, to an operator that acts on the thermofield double Hilbert space $\H_\TFD$.   However, because of fluctuations, $H_r$ and $H_\ell$ do
not separately have  large $N$ limits at $\beta<1/T_\HP$.      For example,  $H_r$ has an expectation value $\la H_r\ra_\beta\sim N^2$ and so does not have
a large $N$ limit.
We can of course subtract this expectation value, but the subtracted operator $H_r-\la H_r\ra_\beta$ has divergent thermal fluctuations $\left\la \left(H_r-\la H_r\ra_\beta\right)^2\right\ra_\beta\sim N^2$ and therefore also does not converge for large $N$ to an operator on $\H_\TFD$.   $\h H$ generates
a group of outer automorphisms of the large $N$ operator algebras of the left and right systems.    This is a group of outer automorphisms because
$H_\ell$ and $H_r$ do not separately have large $N$ limits.

\noindent{{\it Acknowledgments}}
I thank H. Liu, R. Longo, R. Mahajan, J. Maldacena, N. Seiberg, J. Turiaci,  and R. Wald for helpful and informative discussions.  Research supported in part by
 NSF Grant PHY-1911298.

% L. Takhtajan
\bibliographystyle{unsrt}

\end{document}